\documentclass[aps,prd,showpacs,superscriptaddress,onecolumn,raggedfooter,raggedbottom]{revtex4-1}   

\usepackage{amssymb}
\usepackage{amsmath}
\usepackage{amsthm}
\usepackage{graphicx}
\usepackage{subfigure}
\usepackage{color}
\usepackage{mathrsfs} 
\usepackage{paralist}

\usepackage{soul}

\usepackage{hyperref}
\hypersetup{hidelinks}
\usepackage{soul}

\usepackage[capitalize,nameinlink]{cleveref}
\crefname{section}{section}{sections}
\crefname{subsection}{subsection}{subsections}
\Crefname{section}{Section}{Sections}
\Crefname{subsection}{Subsection}{Subsections}
\Crefname{figure}{Figure}{Figures}
\crefformat{equation}{\textup{#2(#1)#3}}
\crefrangeformat{equation}{\textup{#3(#1)#4--#5(#2)#6}}
\crefmultiformat{equation}{\textup{#2(#1)#3}}{ and \textup{#2(#1)#3}}
{, \textup{#2(#1)#3}}{, and \textup{#2(#1)#3}}
\crefrangemultiformat{equation}{\textup{#3(#1)#4--#5(#2)#6}}%
{ and \textup{#3(#1)#4--#5(#2)#6}}{, \textup{#3(#1)#4--#5(#2)#6}}{, and \textup{#3(#1)#4--#5(#2)#6}}
\Crefformat{equation}{#2Equation~\textup{(#1)}#3}
\Crefrangeformat{equation}{Equations~\textup{#3(#1)#4--#5(#2)#6}}
\Crefmultiformat{equation}{Equations~\textup{#2(#1)#3}}{ and \textup{#2(#1)#3}}
{, \textup{#2(#1)#3}}{, and \textup{#2(#1)#3}}
\Crefrangemultiformat{equation}{Equations~\textup{#3(#1)#4--#5(#2)#6}}%
{ and \textup{#3(#1)#4--#5(#2)#6}}{, \textup{#3(#1)#4--#5(#2)#6}}{, and \textup{#3(#1)#4--#5(#2)#6}}
\crefdefaultlabelformat{#2\textup{#1}#3}

\begin{document}

\title{Kink-Antikink Interaction Forces and Bound States in a nonlinear Schr{\"o}dinger Model with Quadratic and Quartic dispersion}

\author{G.~A.\ Tsolias}
\affiliation{Department of Mathematics and Statistics, University of Massachusetts,Amherst, MA 01003-4515, USA}

\author{Robert J.\ Decker}
\affiliation{Mathematics Department, University of Hartford, 200 Bloomfield Ave., West Hartford, CT 06117, USA}

\author{A.\ Demirkaya}
\affiliation{Mathematics Department, University of Hartford, 200 Bloomfield Ave., West Hartford, CT 06117, USA}

\author{T.J. Alexander}
\affiliation{Institute of Photonics and Optical Science (IPOS), School of Physics, The University of Sydney, NSW 2006, Australia}

\author{Ross Parker}
\affiliation{Department of Mathematics, Southern Methodist University, 
Dallas, TX 75275, USA}

\author{P.~G.\ Kevrekidis}
\affiliation{Department of Mathematics and Statistics, University of Massachusetts,Amherst, MA 01003-4515, USA}

\begin{abstract}
In the present work we explore the competition of quadratic and quartic dispersion
in producing kink-like solitary waves in a model of the nonlinear Schr{\"o}dinger
type bearing cubic nonlinearity. We present the first 6 families of multikink solutions
and explore their bifurcations as the strength of the quadratic dispersion is varied. We reveal a rich bifurcation structure
for the system, connecting two-kink states with states involving 4-, as well as
6-kinks. The stability of all of these states is explored. For each family,
we discuss a ``lower branch'' adhering to the energy landscape of the 2-kink
states. We also, however, study in detail the ``upper branches'' bearing 
higher numbers of kinks. In addition to computing the stationary states
and analyzing their stability within the partial differential equation model, we develop an effective 
particle ordinary differential equation theory that is shown to be surprisingly efficient in capturing 
the kink equilibria and normal (as well as unstable) modes. Finally, the
results of the bifurcation analysis are corroborated by means of direct numerical
simulations involving the excitation of the states in a targeted way in order
to explore their instability-induced dynamics.

\end{abstract}

\maketitle

\section{Introduction}

In the study of nonlinear dispersive waves, arguably one of the most
well-established models with a wide range of possible applications
is the nonlinear Schr{\"o}dinger equation~\cite{sulem,ablowitz1,mjarecent}.
Its relevance extends from mean-field limits of atomic 
gases~\cite{stringari,pethick,siambook}, to the propagation of
the envelope of the electric field in optical fibers~\cite{hasegawa,kivshar}
and from water waves~\cite{mjarecent} to plasmas~\cite{plasmas} and beyond.
Nevertheless, recent studies have recognized the experimental relevance
and theoretical interest in exploring realms beyond those of purely
quadratic dispersion, as accompanying the prototypical cubic
nonlinearity (stemming from the Kerr effect~\cite{kivshar} or the 
s-wave scattering of bosons~\cite{stringari,pethick}).

More concretely, over the past few years, a new direction
within nonlinear optics has stemmed from the ability to engineer
dispersion in optical systems in the laboratory, potentially 
completely eliminating quadratic dispersion and enabling quartic
dispersion to be dominant~\cite{pqs}. This has led to the 
experimental observation of the so-called pure-quartic solitons (PQS)~\cite{pqs}
and subsequently the realization of the pure-quartic soliton laser~\cite{pqs3}.
Numerous other possibilities have emerged from this research thread,
including, but not limited to, the ability to program dispersion 
of higher order in fiber lasers~\cite{RungePRR2021}, the possibility
to explore the competing interaction of quadratic and quartic
dispersion for bright~\cite{pqs2,bernd,Parker2021} or dark/kink-like solitary
waves~\cite{OL22,TsoliasJPA2021}, and the study the self-similar
propagation of pulses in the presence of gain~\cite{OL20}
or their nature in the absence of Galilean invariance~\cite{galileo}.
It should be noted that some of these topics (especially with regards to
bright solitary waves) have been the focus of
earlier well-known works~\cite{KarpmanPLA1994,KarpmanPRE96}.
It is also noteworthy that a number of studies have explored
the existence and stability of solutions in related models
bearing 4th order dispersion (or competing dispersions), as well
as their potential for collapse~\cite{beam_demirkaya,atanas,atanas2}.

 In the present work we revisit this interesting class of models,
 aiming to offer a systematic exploration of the branches of 
 kink-like (dark-soliton~\cite{djf}) excitations in the presence
 of quadratic and quartic competing dispersions. A short description
 of the relevant features was presented in~\cite{OL22}. Here, we
 systematically expand upon the branches of solutions noted in this earlier work and
 examine the bifurcation of these solutions in detail. Indeed, we examine the first 6 families
 of states among the ones possible, classified on the basis of the separation between the 
 kink and the antikink. Our emphasis is not on the simpler single
 branch of kink solutions, but rather on the considerably more
 elaborate feature of the quadratic-quartic model, namely the 
 possibility of existence of multi-kink bound states. We start
 from the simpler 2-kink states, which form the so-called ``lower
 branches'' of our bifurcation diagrams and continue the solutions
 in one of the key parameters of the system, namely the strength of the quadratic dispersion.
For all of the relevant families (except for the ``exceptional''
0th family which seems to emanate from the small amplitude limit),
the branches feature a characteristic turning point which leads to
an ``upper branch'' of states. The latter nucleates either one
or two pairs of additional kinks, leading to states involving
4-kink and 6-kink solutions. We identify all of these states systematically
and present a comprehensive overview of their stability properties.
Equally importantly, in the limit of large $\beta_2$ (the quadratic dispersion parameter),
we develop a theoretical formulation of the interacting kinks as ``effective
particles'' (see, e.g., also our earlier considerations in~\cite{TsoliasJPA2021}). This, in turn, allows us to identify the equilibrium configurations and their kink locations in the resulting interacting particle system, and examine the linear properties of these particles around the equilibria.  We find that this particle picture is remarkably accurate at 
capturing the unstable and stable modes of the multi-kink states.
Whenever relevant, we also complement the existence and stability studies
with dynamical computations exploring the fate of the unstable states.

Our presentation will be structured as follows. In Section II, we will
provide the general theoretical framework of the problem and the methodology
for exploring the multi-kink interacting particle system. In Section III, 
 the center-theme of our work, we provide the existence and stability analysis, and compare our theoretical results with those obtained from dynamical simulations.  Finally, in Section IV, we summarize
 our findings and offer some directions for further study.
 In the Appendix, some of the details of our numerical computations
 are systematically provided for the reader that is more keenly interested
 in the practical aspects of the numerical methods.
  
\section{Model Setup and Analysis}

The generalized variant of the nonlinear Schr{\"o}dinger (GNLS) equation 
that we study in the present work is~\cite{OL22}:
\begin{equation}
iu_t+\frac{\beta_4}{4!}u_{xxxx}-\frac{\beta_2}{2}u_{xx}+\gamma |u|^2u=0,
\label{NLS}
\end{equation}
where in an optical context $\beta_4$ characterizes the strength of the fourth-order dispersion, $\beta_2$ the strength of the second-order dispersion and $\gamma$ the strength of the cubic nonlinearity.  We focus on the so-called quartic normal dispersion regime, where stable dark solitons have been found~\cite{OL22} in the presence of attractive nonlinearity, and so take $\beta_4 > 0$ and $\gamma > 0$. Later we will restrict to $\beta_4=1$ and $\gamma=1$.

{\subsection{Stationary States and Spectral Stability}}

Eq. \cref{NLS} is a Hamiltonian system, with conserved energy $\mathcal{E}$ given by
\begin{equation}\label{eq:E}
\mathcal{E}(u) = \frac{1}{2} \int_{-\infty}^\infty \left( \frac{\beta_4}{4!}|u_{xx}|^2 + \frac{\beta_2}{2}|u_{x}|^2 + \frac{\gamma}{2} |u|^4 \right) dx.
\end{equation}
Separating real and imaginary parts by taking $u = u_R + i u_I$, Eq. \cref{NLS} can be written in standard Hamiltonian form as 
\begin{equation}\label{eq:NLSHam}
\frac{\partial u}{\partial t} = J \mathcal{E}'(u(t)),
\end{equation}
where $u = (u_R, u_I)^T$, $\mathcal{E}'(u(t))$ is the functional derivative of $\mathcal{E}(u)$ evaluated at $u(t)$ and $J$ is the standard symplectic matrix
\begin{equation}\label{eq:J}
J = \begin{bmatrix}
0 & 1 \\ -1 & 0
\end{bmatrix}.
\end{equation}
Eq. \cref{eq:NLSHam} then becomes the pair of real-valued equations
\begin{eqnarray}\label{eq:NLScomponent}
(u_R)_t=-\frac{\beta_4}{24} (u_I)_{xxxx}+\frac{\beta_2}{2} (u_I)_{xx}-({u_R}^2+{u_I}^2)u_I \\
(u_I)_t=\frac{\beta_4}{24} (u_R)_{xxxx}-\frac{\beta_2}{2} (u_R)_{xx}+({u_R}^2+{u_I}^2)u_R.
\end{eqnarray}
In what follows, we are interested in stationary (time-independent amplitude) solutions of the form:
$u(x,t)=e^{i\mu t}\phi(x)$. Substituting this ansatz into \cref{NLS}, we get
the steady state model for the amplitude $\phi$
\begin{equation}\label{eq:NLSmu_main}
\frac{\beta_4}{4!}\phi''''-\frac{\beta_2}{2}\phi''-\mu \phi+\gamma\phi^3=0.
\end{equation}
In addition to the solution $\phi=0$, Eq. (\ref{eq:NLSmu_main}) has continuous wave (CW) solutions 
$\phi = \pm \sqrt{\mu/\gamma}$. 

To facilitate our understanding of the nature of these fixed points, we rewrite \cref{eq:NLSmu_main} as a system of four first-order ordinary differential equations, using $U = (u_1, u_2, u_3, u_4) = (\phi, \phi', \phi'', \frac{\beta_4}{24} \phi''')$.  Eq. \cref{eq:NLSmu_main} then becomes the first order system in $\mathbb{R}^4$
\begin{equation}\label{eq:Fsystem_main}
U' = F(U) = \begin{bmatrix}
u_2 \\ u_3 \\ \frac{24}{\beta_4} u_4 \\ \frac{\beta_2}{2} u_3 + \mu u_1 - \gamma u_1^3
\end{bmatrix}.
\end{equation}
For $\beta_4 > 0$, $\mu >0$, and all $\beta_2$, the linearization about $\phi=0$ has a pair of real eigenvalues $\pm \alpha$ and a pair of imaginary eigenvalues $\pm \beta i$, thus the corresponding equilibrium of \cref{eq:Fsystem_main} has a two-dimensional center subspace, and $\phi = 0$ is a saddle-center fixed point. The eigenvalues of the linearization about the CW states $\phi=\pm \sqrt{\mu/\gamma}$ are instead
\[
\lambda = \pm \sqrt{ \frac{6 \beta_2 \pm 2 \sqrt{9 \beta_2^2 - 12 \beta_4 \mu}}{\beta_4} }.
\]
For fixed $\beta_4>0$ and $\mu>0$, corresponding equilibria $S^\pm = (\pm \sqrt{\mu/\gamma}, 0, 0, 0)$ are saddle points of \cref{eq:Fsystem_main} when $\beta_2 > -\beta_2^*$, where
\begin{equation}\label{eq:beta2star}
\beta_2^* = 2\sqrt{\frac{\beta_4 \mu}{3}}.
\end{equation}
The stable and unstable manifolds of $S^\pm$ are both two-dimensional. When $|\beta_2| < \beta_2^*$, the spatial eigenvalues are a complex quartet $\pm a \pm b i$, topologically corresponding to a saddle-spiral, and when $|\beta_2| > \beta_2^*$, they are two pairs of real eigenvalues $\pm b_1$ and $\pm b_2$, leading to a saddle point in the four dimensional space.

In our computations that will follow, the conditions $-\beta_2^* < \beta_2 < \beta_2^*$
will play a pivotal
role providing a set of bounds for $\beta_2$ under which the kink-antikink states
of interest will exist. Later we will restrict our attention to the specific case of $\mu=5$.

A kink $\phi_k$ is a solution to \cref{eq:NLSmu_main} connecting the CW state at $-\sqrt{\mu/\gamma}$ ($S^-$) to the one at $\sqrt{\mu/\gamma}$ ($S^+$). From a spatial dynamics perspective, this is a heteroclinic orbit connecting the saddle points $S^-$ and $S^+$. If $\phi$ is a solution to \cref{eq:NLSmu_main}, so is $-\phi$, thus for every kink solution $\phi_k$ there is a corresponding anti-kink $-\phi_k$. When $\beta_4 = 0$ and $\beta_2 > 0$, the exact formula for the stationary kink is given by
\begin{equation}\label{eq:NLS2kink}
\phi_k(x) = \sqrt{\frac{\mu}{\gamma}}\tanh\left( \sqrt{\frac{\mu}{\beta_2}} x \right).
\end{equation}
We take the existence of a primary kink solution to \cref{eq:NLSmu_main} as a hypothesis in what follows.

To study the stability of these solutions, we consider the linearization around $u(x,t)=e^{i\mu t}\phi(x)$, where $\phi(x)$ is a solution to \cref{eq:NLSmu_main}. Adding the perturbation as follows $u(x,t)=e^{i\mu t}[\phi(x)+v(x,t)]$, where $v(x,t)=v_{R}+iv_{I}$, substituting it into \cref{NLS}, 
we obtain two equations:
\begin{align*}
(v_{R})_t&=-\frac{\beta_4}{4!}v_{I}''''+\frac{\beta_2}{2}v_{I}''+\mu v_{I}-\gamma \phi^2v_{I}\\
(v_{I})_t&=\frac{\beta_4}{4!}v_{R}''''-\frac{\beta_2}{2}v_{R}''-\mu v_{R}+3\gamma \phi^2v_{R}
\end{align*}
which can be written as
\begin{equation}
\label{eq:stab}
\frac{\partial }{\partial t }\left[ 
\begin{array}{c}
v_{R}  \\ 
v_{I}
\end{array}%
\right] =\left[ 
\begin{array}{cc}
0 & -\mathcal{L}_{-}(\phi) \\ 
\mathcal{L}_{+}(\phi) & 0
\end{array}
\right]  \left[ 
\begin{array}{c}
v_{R}  \\ 
v_{I}
\end{array}%
\right]
= -J \mathcal{L}(\phi) \left[ 
\begin{array}{c}
v_{R}  \\ 
v_{I}
\end{array}%
\right]
, 
\end{equation}
where 
\begin{align*}
\mathcal{L}_{+}(\phi) &=\frac{\beta_4}{4!}D^4-\frac{\beta_2}{2}D^2-\mu+3\gamma \phi^2\\
\mathcal{L}_{-}(\phi) &=\frac{\beta_4}{4!}D^4-\frac{\beta_2}{2}D^2-\mu+\gamma \phi^2\\
\mathcal{L}(\phi) &= \left[ 
\begin{array}{cc}
\mathcal{L}_{+}(\phi) & 0 \\ 
0 & \mathcal{L}_{-}(\phi) 
\end{array}
\right].
\end{align*}
In what follows we find the lowest six families (as defined in Section \ref{numerical})  of stationary kink-antikink solutions numerically and examine their stability using Eq.~\cref{eq:stab}.

\subsection{{Effective Particle Model}}
\label{eff_part}

Using a method due to N. Manton~\cite{manton}, we now derive an ODE model of kink - antikink interaction. For our model the Lagrangian is
\begin{align}
    L = \int_{-\infty}^{\infty} \mathcal{L} dx = \int_{-\infty}^{\infty} \left(\frac{i}{2}(u^*_tu - u^*u_t) -\frac{\beta_2}{2}u^*_x u_x - \frac{\beta_4}{4!} u^*_{xx}u_{xx} 
    -\mathbin{\frac 1 2 \gamma |u|^4} 
    \right) dx.
\end{align}
Invariance under translations gives rise to the conserved quantity 
\begin{align}
    P = \int^\infty_{-\infty}  \frac i 2 \left( u^*_xu - u^*u_x \right) dx
\end{align}
which is the total momentum \(P\) of the field \(u\). In order to calculate the force \(F\) between a kink and an antikink, we consider the momentum included in a finite interval \([x_1,x_2]\) and we differentiate with respect to time \(t\). 
\begin{align}\nonumber
    F = \frac{dP}{dt} &= \int^{x_2}_{x_1} \frac i 2 \left( u^*_{xt} u +u^*_x u_{t} - u^*_{t} u_x- u^* u_{xt} \right) dx\\\nonumber
    &= \int^{x_2}_{x_1} i \left( u^*_x u_t - u^*_t u_x \right) dx + \frac i 2 \left[ u^*_t u - u^*u_t  \right]^{x_2}_{x_1}\\
    &=  \left[\frac{\beta_2} 2 u^*_x u_{x} -\frac{\beta_4}{4!} (u^*_x u_{xxx} 
    -u^*_{xx} u_{xx} +u^*_{xxx}u_x) - \frac{\gamma}{2} |u|^4 + \frac i 2  (u^*_t u - u^* u_t  ) \right]^{x_2}_{x_1}
\end{align}
For \( u(t,x) = e^{i \mu t} \phi(x)\), where \(\phi(x)\) is a {real }static field this expression simplifies to 
\begin{align}\label{eq:Force12}
    F =  \left[\frac{\beta_2} 2 {\phi'}^2 +\frac{\beta_4}{4!} \left( {\phi''}^2 - 2 \phi' \phi''' \right)
     - \frac{\gamma}{2} \phi^4  + \mu \phi^2 \right]^{x_2}_{x_1} = F_{x_2} - F_{x_1}
\end{align}
which is zero, as expected for a static solution (the quantity inside the brackets is constant if \(\phi\) satisfies Eq.~(\ref{eq:NLSmu_main}).

Now, suppose we have a superposition of a kink centered at \(x=-X\) and an antikink centered at \(x=X\).  Then the force on the antikink due to the kink is given by Eq.~(\ref{eq:Force12}) for \(x_1=0\) and \(x_2\rightarrow\infty\), {i.e., integrating across the antikink to find the force exerted on it due to the change of its momentum.}

For \(x_2\rightarrow \infty\), let \(\phi\rightarrow {-\sqrt{\mu / \gamma}} \).
Then,
\begin{align}
    F_{x_2} = {\frac{\mu^2}{2\gamma}}
\end{align}
For \(x\) in the region between the two kinks, let 
\(\phi(x) = {\sqrt{\mu / \gamma}} -\eta (x)\),
for \(\eta\) small. Then keeping up to second order terms we get
\begin{align}
F_{x_1}
    &\approx\left[\frac{\beta_2} 2 {\eta'}^2 +\frac{\beta_4}{4!} \left( {\eta''}^2  - 2 \eta' \eta''' \right) +\frac{\mu^2}{2\gamma} -2 \mu \eta^2   \right]_{x=0}
\end{align}
Therefore the force acting on the antikink is given  in terms of \(\eta(x)\) by the following expression:
\begin{align}\label{eq:ForceEta}
    F  \approx \left[ -\frac{\beta_2} 2  {\eta'}^2 -\frac{\beta_4}{4!} {\eta''}^2 
    +\frac{\beta_4}{4!} 2\eta' \eta'''
    + 2 \mu{\eta}^2 \right]_{x=0}
\end{align}

If \(X\) is large enough, so that the two kinks are well separated, then \(\eta(x)\) can be very well approximated by the superposition of their tails. In particular, for large positive \(x\), a single kink can be written as \(\phi_K (x) = \sqrt{\frac{\mu}{\gamma}} -\chi(x)\) (and similarly, for large negative \(x\)  a single antikink can be written as \(\phi_{AK} (x) = \sqrt{\frac{\mu}{\gamma}} -\chi(-x)\)), where the tail \(\chi(x)\) satisfies the linearized problem
\begin{align}
     2 \mu\chi - \frac{\beta_2} 2 \chi'' + \frac{\beta_4}{4!} \chi''''  = 0
\end{align}
For \(\beta_2 <2\sqrt{\frac{\beta_4 \mu}3}\) the linearized equation has vanishing solutions of the form 
 \begin{align}\label{eq:tail}
\chi(x) =  e^ {-r x} (A  \cos(k x ) + B  \cos(k x )  )
\end{align}  
where 
\begin{align}
    r = \sqrt{\frac{2\sqrt{3\beta_4\mu}+3\beta_2}{\beta_4}} && \text{ and } & & k = \sqrt{\frac{2\sqrt{3\beta_4\mu}-3\beta_2}{\beta_4}}.
\end{align}
Then, the superposition of the tails gives
\begin{align}
    \sqrt{\frac{\mu}{\gamma}} -\eta(x) {= \phi(x)} & \approx \phi_K(x+X) + \phi_{AK}(x-X) -\sqrt{\frac{\mu}{\gamma}}, 
\end{align}
from which we obtain
\begin{align}
    \eta(x) {\approx} \chi (x+X) + \chi (-x+X)
\end{align}
Substituting this into Eq.~(\ref{eq:ForceEta}) and using Eq.~(\ref{eq:tail}), we finally get
\begin{align}
    F \approx e^ {-2r X} \left(\left( 2(A^2 - B^2) \frac{r^2 k^2 \beta_4} 3 + 4 A B r k \beta_2 \right) \cos(2 k X) + \left(  4 A B \frac{r^2 k^2 \beta_4} 3 -2(A^2 - B^2) r k \beta_2 \right)  \sin(2 k X)\right)
\end{align}

For an effective ODE description we need to find the inertial mass of the kinks. For \(c\) small enough, 
our numerical computations of traveling kinks
suggest that the field \(u\)  configuration 
can be written as
\(u(x,t) = e^{i\mu t} \left(\phi_K (x-ct) - c^2 v(x-ct) + i c\, w(x-ct) \right). \)
This corresponds to the profile of a kink \(\phi_K\) moving to the right with constant speed \(c\), since \( |u|^2 = \phi_K^2(x-ct) + \mathcal{O}({c^2})\),
while $w$ denotes to the leading order imaginary
(linear in $c$) and $v$ the leading order real
(quadratic in $c$) correction.

Now, from Eq.~(\ref{NLS}) we get
\begin{align}
     i (u^*_t u + u^* u_t) &= 
    - \frac{\beta_2} 2 (u^*_{xx} u-u^*u_{xx}) 
    +\frac{\beta_4}{4!} (u^*_{xxxx}u - u^*u_{xxxx})
\end{align}
and since \(|u|^2\) is a function of \( x-ct\), the time derivative can be expressed as a spatial derivative multiplied by \(-c\).
{
\begin{align}
    i (u^*_t u + u^* u_t) = i \left(|u|^2\right)_t =  -ic \left(|u|^2\right)_x 
\end{align}
}

Integrating over \(x\) gives
\begin{align}
    -ic |u|^2 &= 
    - \frac{\beta_2} 2 (u^*_x u-u^*u_x) 
    +\frac{\beta_4}{4!} (u^*_{xxx}u - u^*_{xx} u_{x} + u^*_{x}u_{xx}-u^*u_{xxx}) -i cK
\end{align}
where \(K\) is an integrating real constant. Of course, for this equation to hold as \(x\rightarrow \infty\), we need \(K = \frac{\mu}{\gamma}\). Integrating one more time over the whole \(x\)-axis and rearranging the terms, we get:
\begin{align}
    \int_{-\infty}^{\infty} \frac{\beta_2} 2 (u^*_x u-u^*u_x) dx &= 
    -i c \int_{-\infty}^{\infty} \left(  \frac \mu \gamma - |u|^2\right) dx
    + \frac{\beta_4}{4!} \int_{-\infty}^{\infty}  (u^*_{xxx}u - u^*_{xx} u_{x} + u^*_{x}u_{xx}-u^*u_{xxx}) dx 
\end{align}
{Then the total momentum of the traveling kink is given by }
\begin{align}
    P &=     \frac{c}{\beta_2}\int_{-\infty}^{\infty} \left( \frac \mu \gamma-|u|^2 \right) dx     + \frac{ \beta_4 c}{12\beta_2} \int_{-\infty}^{\infty}   \left({\phi_K} w''' -{\phi_K}'w'' + {\phi_K}'' w' - {\phi_K}''' w\right) dx +\mathcal{O}(c^2)
\end{align}
{and using the definition \(P=Mc\) for the inertial mass we find }
\begin{align}
    M  &=     \frac{1}{\beta_2}\int_{-\infty}^{\infty} \left( \frac \mu \gamma-\phi_K^2 \right) dx     + \frac{ \beta_4 }{12\beta_2} \int_{-\infty}^{\infty}  \left({\phi_K} w''' -{\phi_K}'w'' + {\phi_K}'' w' - {\phi_K}''' w\right) dx
\label{Mass_equation}
\end{align}
where {\(w\) must satisfy}
\begin{align}
    -\mu w - \frac{\beta_2}{2}w'' + \frac{\beta_4}{4!} w'''' + \gamma \phi_K^2 w = {\phi_K}'
\label{w_equation}   
\end{align}
In what follows, we will consider the mass that solely stems from the first term of Eq. (\ref{Mass_equation}),
i.e., the standard renormalized mass of the defocusing NLS problem (that has
also been used, e.g., towards the proof of the stability of the dark
solitons thereof in, e.g.,~\cite{igorb}. This will be justified a posteriori
via the comparison of our results with the detailed numerical computations.
A rigorous justification of this choice from first principles is an interesting
topic for future study.

\section{Numerical Results and Comparison}
\label{numerical}

We now restrict our attention to numerical solutions of Eq. (\ref{eq:NLSmu_main}) with $\mu=5$, $\gamma=1$, and $\beta_4=1$. Letting $\beta_2$ vary, we get families of solutions to Eq. (\ref{eq:NLSmu_main}), where the members of each family are connected by numerical continuation with respect to $\beta_2$. We start by finding kink-antikink solutions at $\beta_2=0$ corresponding to larger and larger separation of the kink and antikink, as in \cite{OL22}. We then use numerical continuation to create families of solutions which we refer to as family 0 (continuation of the smallest possible separation of kink-antikink at $\beta_2=0$), family 1 (continuation of the second smallest possible separation of kink-antikink at $\beta_2=0$), and so on for families 2, 3, 4, and 5.  
Additional families have been identified in our numerical computations, however
to keep the presentation more succinct, we do not discuss them here.

\begin{figure}[tbp]
\includegraphics[scale=0.36]{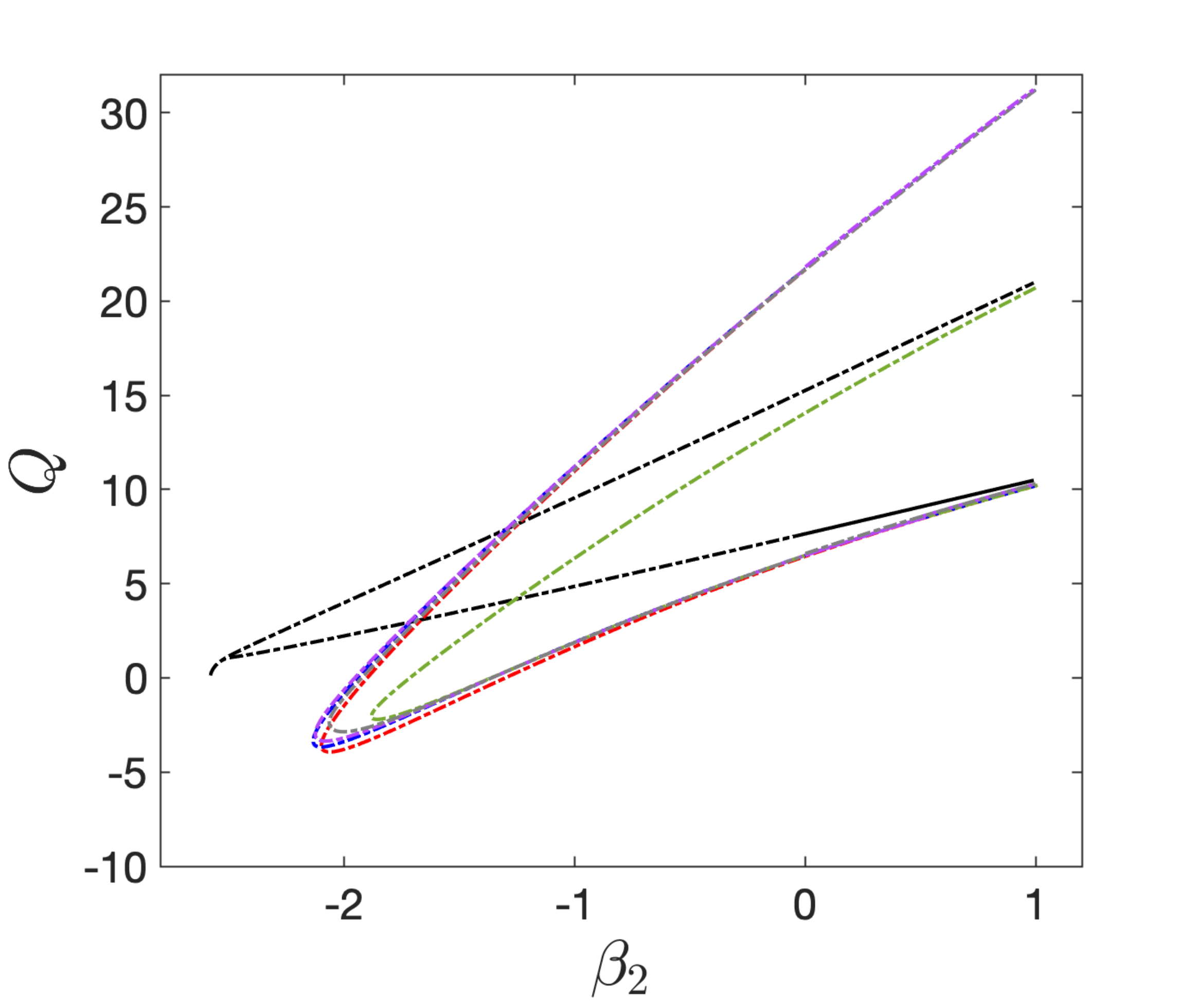}
\caption{ Bifurcation diagrams of six families of solutions: family 0 (black), family 1 (red), family 2 (blue), family 3 (green), family 4 (purple), family 5 (gray). In all cases $\mu=5$, $\gamma=1$, and $\beta_4=1$. }
\label{fig:bif_plots}
\end{figure}

An overarching summary of our results is shown in Fig.~\ref{fig:bif_plots}. Here we see that each family of solutions has both an upper and lower branch, which are connected by numerical continuation, with the exception of family 0, where the upper branch was not created as a numerical continuation of the lower branch. 
Before we delve into the details of individual branches, we identify the main features encompassing all the branches on the figure.  Throughout our analysis we make use of the complementary power $Q$ to characterize the families of solutions,
\begin{equation}
\label{eq:Q}
Q = \int_{-L}^{L} \left(\frac{\mu}{\gamma} - \phi^2\right) dx,
\end{equation}
where $L$ is the half-width of the stationary solution domain. This is effectively the same
quantity as the one defined by $M$ in the previous section, however to more clearly
distinguish between the two (bearing in mind that in the latter there is, in principle,
also a contribution $\propto \beta_4$, we use a different symbolism here.

We can see on the right side of Fig.~\ref{fig:bif_plots}, as $\beta_2 \rightarrow 1$, that the branches
form 3 groups, with each group distinguished by the number of kinks present (2, 4 and 6 in increasing complementary power). The lowermost group involves what we will
hereafter term ``lower branches'' for all the families considered below.
This concerns the states involving 2 kinks that are well-separated in this
large and positive $\beta_2$ limit. The next group, encompassing
solely family 0 (black upper branch) and family 3 (green upper branch),
involves solutions consisting of 4 kinks. Finally, after a similar
``jump'' in complementary power, we encounter all remaining
upper branches (e.g., the red of family 1, the blue of family 2,
the purple of family 4 and the gray of family 5), which are all solutions with 6 kinks. These results relate to the large and positive $\beta_2$ limit, but in the case where
$\beta_2$ becomes negative, we encounter a similar partition between
the branches. Namely, the black and green branches are somewhat ``special''.
The former tends to a limit of 
progressively smaller complementary power, i.e., tending to a small-amplitude steady oscillation about the CW solution itself, while the latter has a turning point 
$\beta_2^{cr}$ which is
distinct from that of all other branches. However the 4 remaining branches
(1, 2, 4 and 5) seem quite similar at the level of this complementary power diagnostic. For these 4 branches all solutions starting on a lower branch feature a turning point for a negative value
of $\beta_2$ (near $-2$) and subsequently continue along an upper
branch towards the 6-kink configuration discussed above. A final observation that we make based on earlier analysis \cite{OL22} is that the CW is modulationally stable whenever $\beta_2 \geq 0$, but when $\beta_2 < 0$ there is a continuous band of modulationally unstable wavenumbers with bounds $k = \pm\sqrt{-12\beta_2/\beta_4}$.
Indeed, the relevant branch(es) can never be stable
for $\beta_2<0$. 
We now turn
to details of each of the relevant families.

\subsection{Family 0}

\begin{figure}[tbp]
\begin{center}
\subfigure[]{
\includegraphics[scale=0.37]{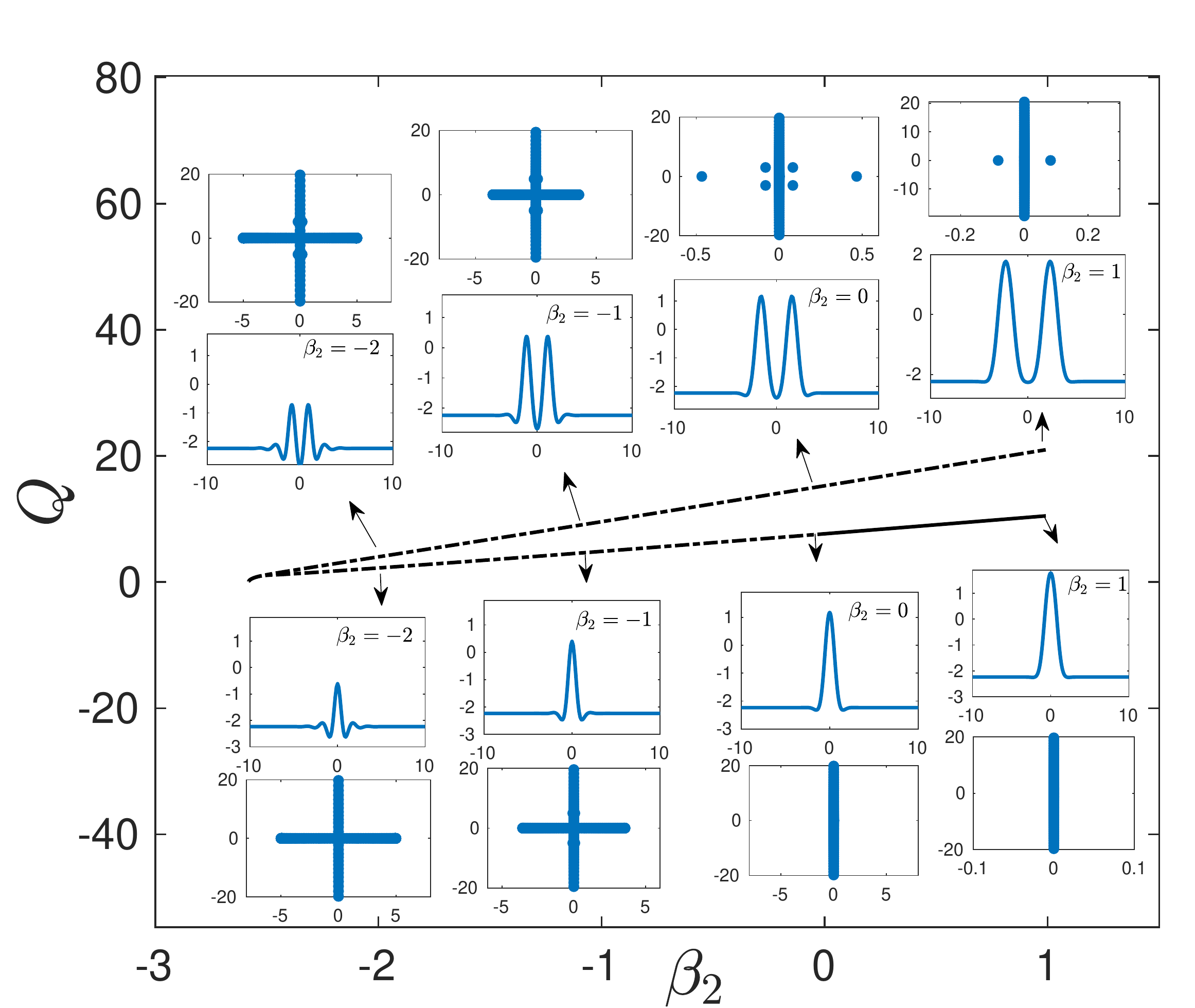}}
\subfigure[  ]{
\includegraphics[scale=0.37]{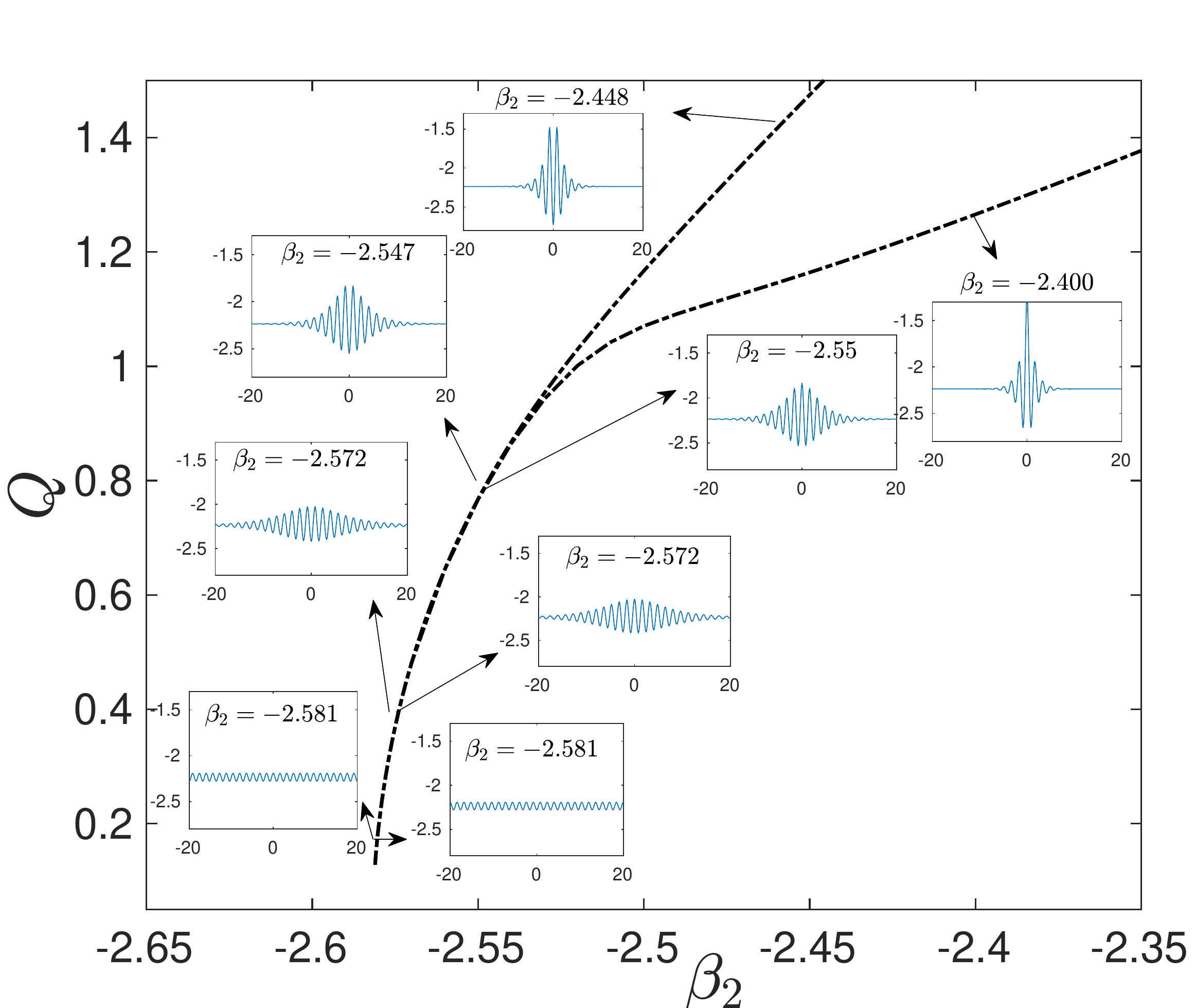}}
\end{center}
\caption{(a) Bifurcation diagram ($Q$ vs $\beta_2$), the corresponding steady state solutions and spectra for Family 0, presented for a sequence of values of
the quadratic dispersion parameter $\beta_2$. (b) the same bifurcation diagram as (a) but zoomed in about the intersection of the upper and the lower curves. 
  }
\label{fig:bif_plots_0}
\end{figure}

\begin{figure}[tbp]
\begin{center}
\subfigure[]{
\includegraphics[scale=0.24]{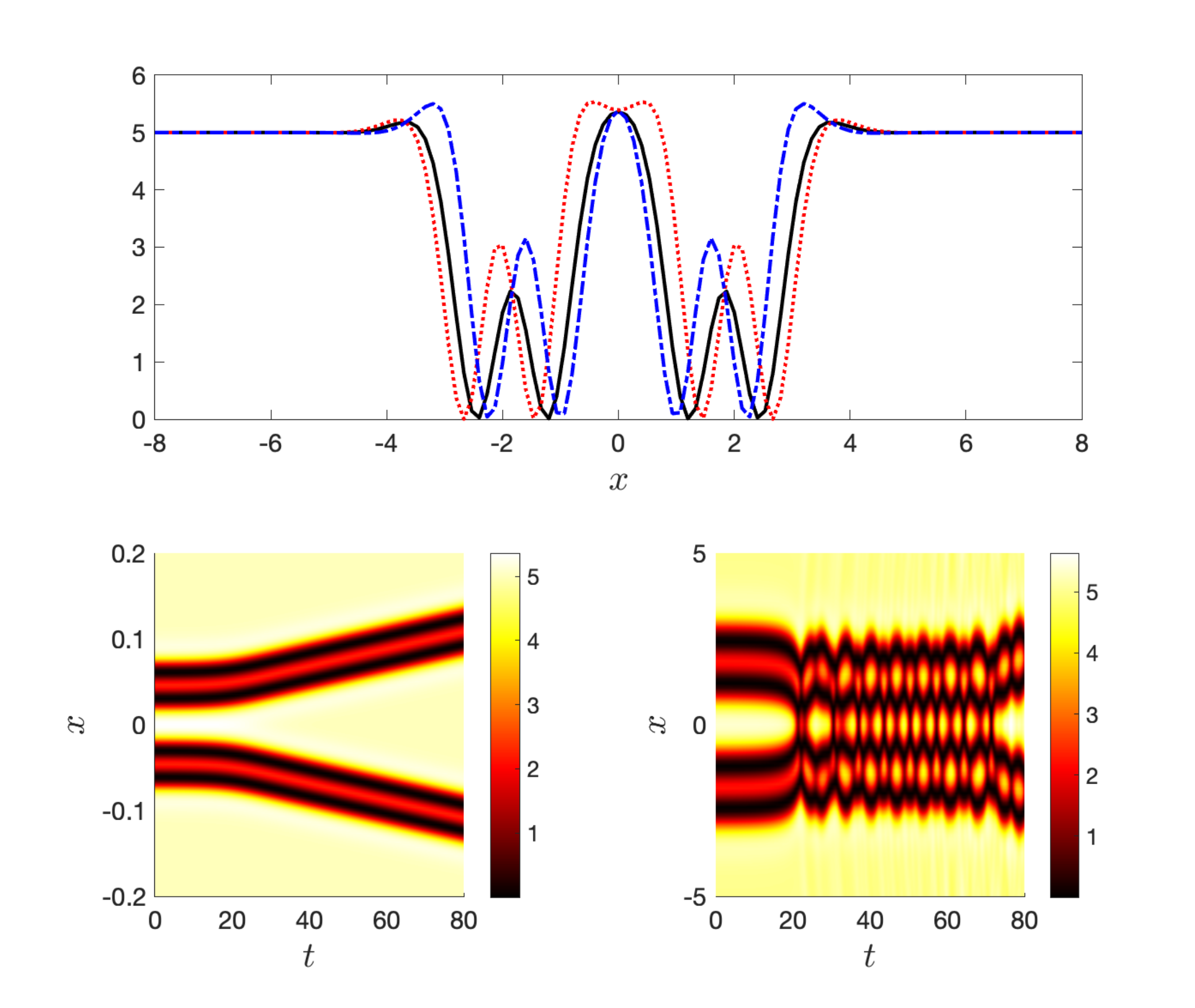}}
\subfigure[]{
\includegraphics[scale=0.28]{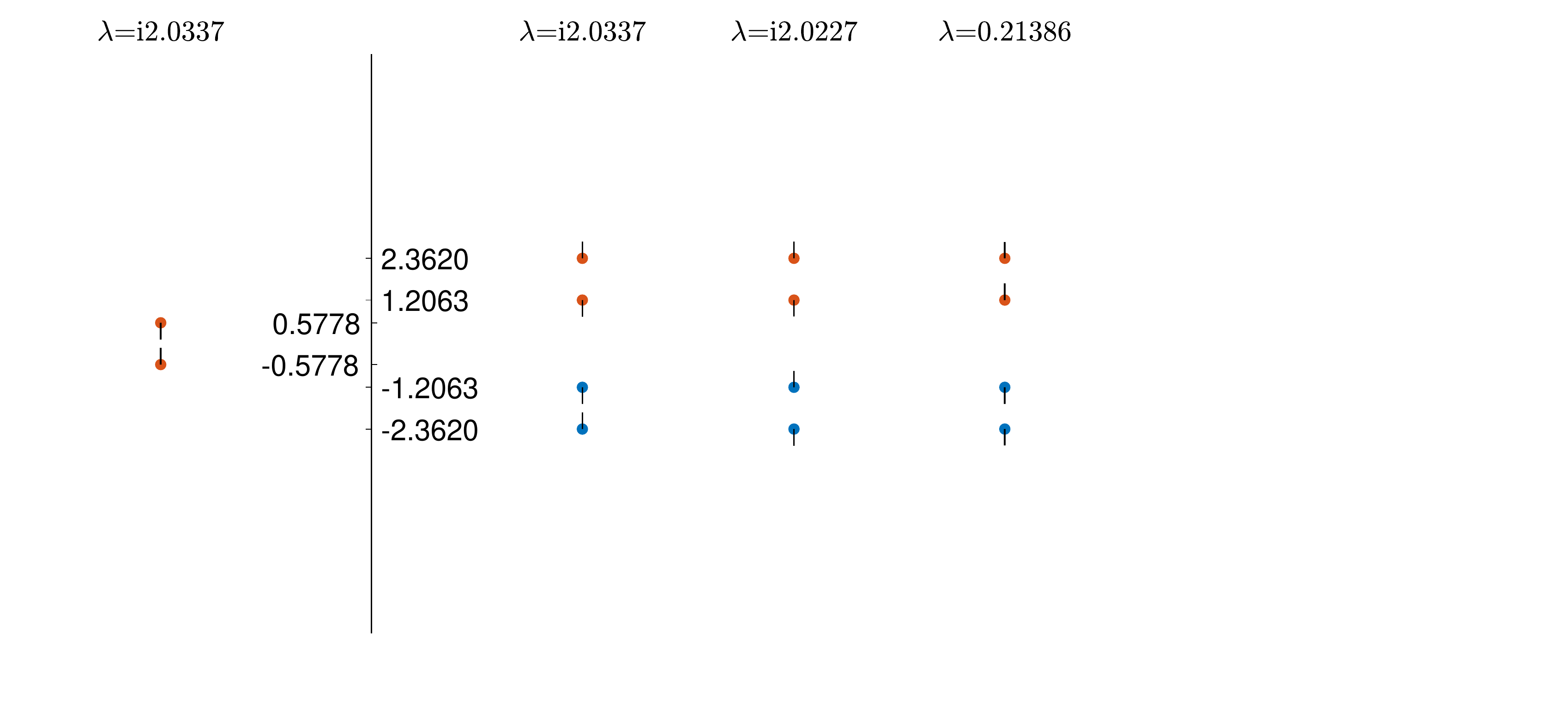}}
\end{center}
\caption{PDE and ODE initial conditions and dynamics for family 0 (upper branch only for PDE). The upper figure of (a) shows the plots of $\|u\|^2$ (black), $\|u+v_1\|^2$ (red), and $\|u-v_1\|^2$ (blue) for $\beta_2=0.5$, where $v_1$ is the eigenfunction corresponding to the only real PDE eigenvalue of 0.2231. The lower left panel of (a) is the contour plot that results from using $\|u+0.01v_1\|^2$ as the initial condition, and the lower right panel of (a) is the contour plot that results from using $\|u-0.01v_1\|^2$ as the initial condition. (b) gives the ODE values for the soliton positions (left of vertical line is
the lower branch, right of the vertical line is the upper branch) and the ODE eigenvalues along the top (again lower branch left and upper branch right). Arrows on the points indicate the initial directions of the solitons (all directions would be reversed if $v_1$ is replaced by $-v_1$.)} 
  
\label{fig:dyn_plots_0}
\end{figure}

We start by describing the family 0 presented in the bifurcation diagram of 
Fig.~\ref{fig:bif_plots_0} and the dynamics plots of Fig.~\ref{fig:dyn_plots_0}.
Fig.~\ref{fig:bif_plots_0}(a) illustrates the profiles and spectral planes
of the stability analysis associated with this branch for different values of $\beta_2$.
If we look at the zoomed in bifurcation diagram on the right, Fig.~\ref{fig:bif_plots_0}(b), we see that the states appear to bifurcate from the CW background as localized wavepackets (with increasing $\beta_2$), for both the lower and upper branches.  These branches tend to a small oscillation about the fixed point $\sqrt{\mu/\gamma}$ (for decreasing $\beta_2$), of approximately 0.04, as $\beta_2$ tends to $\beta_2^* \approx -2.58$. This limiting amplitude gets smaller as the simulated $x$-domain gets larger (for double the size of the $x$-domain this amplitude is approximately $0.02$) and so it is reasonable to assume that the limiting steady state flattens out as the size of the $x$-domain approaches infinity. 

Following the lower branch to increasing $\beta_2$ (lower panels of Fig.~\ref{fig:bif_plots_0}(a)) we see that the state can be described as a kink-anti-kink pair with an increasing separation distance.  Indeed, the analysis of~\cite{TsoliasJPA2021} which is generalized
in Sec.~\ref{eff_part}, shows that the quartic dispersion induces an oscillatory
tail in the kinks which (competing with the quadratic
dispersion), in turn, enables the possibility of 
bound states between two kink-like structures.
This landscape consists of an alternation of
local energy minima (such as the present one), forming center points in the 
landscape of the soliton center dynamics,
and local energy maxima, which, naturally, 
correspond to saddle points in the relevant landscape.
For the centers like the one corresponding to the lower bifurcation branch, we expect
stability, at least as far as the motion
of the kink centers is concerned, and indeed we see in the bottom right of Fig.~\ref{fig:bif_plots_0}(a) that there are no instability eigenvalues (when $\beta_2 > 0$).

Turning our attention now to the upper branch in Fig.~\ref{fig:bif_plots_0} we see that this corresponds to a state consisting of four kinks (i.e., four zeros in the amplitude at large $\beta_2$), but it also bifurcates from the nearly flat state at $\beta_2^*\approx -2.58$.  Indeed, it appears (see the bottom insets
of Fig.~\ref{fig:bif_plots_0}(b)) that
the two and 4 kink states merge, in the sense of the complementary power Q, in the small amplitude limit (about the CW).  In contrast to the lower branch however, the upper branch appears to always be unstable. Recall that for this family, the upper branch is not a numerical continuation of the lower branch (as is the case for the other families) but rather is calculated the same way that the lower branch is calculated, using a numerical continuation from a carefully chosen initial steady state at $\beta_2=0$.

While an oscillatory eigenvalue quartet seems to exist,
we will not discuss such instabilities at length,
as they appear in our computations to be strongly
dependent on the computational domain size.
Indeed, similarly to other such examples in the
realm of dark solitons (starting with the work
of~\cite{johkiv}), the presence of so-called anomalous
modes, pertaining to the motion of the solitary waves, inside the continuous spectrum gives rise to such resonances which are
domain-size dependent as the latter determines the 
(finite-domain-induced) ``quantization''
of the continuous spectrum. On the other hand,
we observe that the 4 kinks have 3 internal
modes in their dynamics (in addition to their 
translational motion which is neutral and pertains
to a so-called Nambu-Goldstone mode associated with
the corresponding invariance). Indeed, for the 2 kinks,
there is only one motion, in addition to their neutral
translation, namely the out-of-phase relative motion thereof,
while generally for $N$ kinks, we should expect $N-1$ such
internal modes associated with the kink relative motions.
In the case of the upper branch of family 0, 
Fig.~\ref{fig:dyn_plots_0} elucidates the situation.
In particular, its right panel uses the approach pioneered by Manton~\cite{manton}
and analyzed above in Sec.~\ref{eff_part} 
to identify the equilibrium kink configuration and performs
a linearization analysis around it, in the form of a $ 4 \times 4$ system
to obtain the effective particle normal modes 
Two of these are oscillatory (featuring one pair moving
towards each other and one away from each other), while the
third is an unstable real mode with the kinks moving in 
opposite outward directions. The kinks centered in the
positive half-line move in unison and so do the ones
in the negative half-line, but these two pairs move in opposite
directions between them. This instability can lead to a 
splitting of the 4-kink bound state into two 2-kink bound states
as shown in Fig.~\ref{fig:dyn_plots_0}(a), but it can also
lead all 4 to collide at the center, featuring a long-lived breathing
state before eventually separating.
Notice that, as explained in the inset, in each
of these dynamical evolution cases, we will
present the case example where we have added the unstable
eigenvector to the configuration, and also the one where
we have subtracted it. These two possibilities have
been used in order to seed the instability in two opposite
directions, as seen in panel (a) of the figure and, similarly,
in other examples of such seeding presented below.

\subsection{Families 1, 3, 5}

\begin{figure}[tbp]
\begin{center}
\subfigure[]{
\includegraphics[scale=0.37]{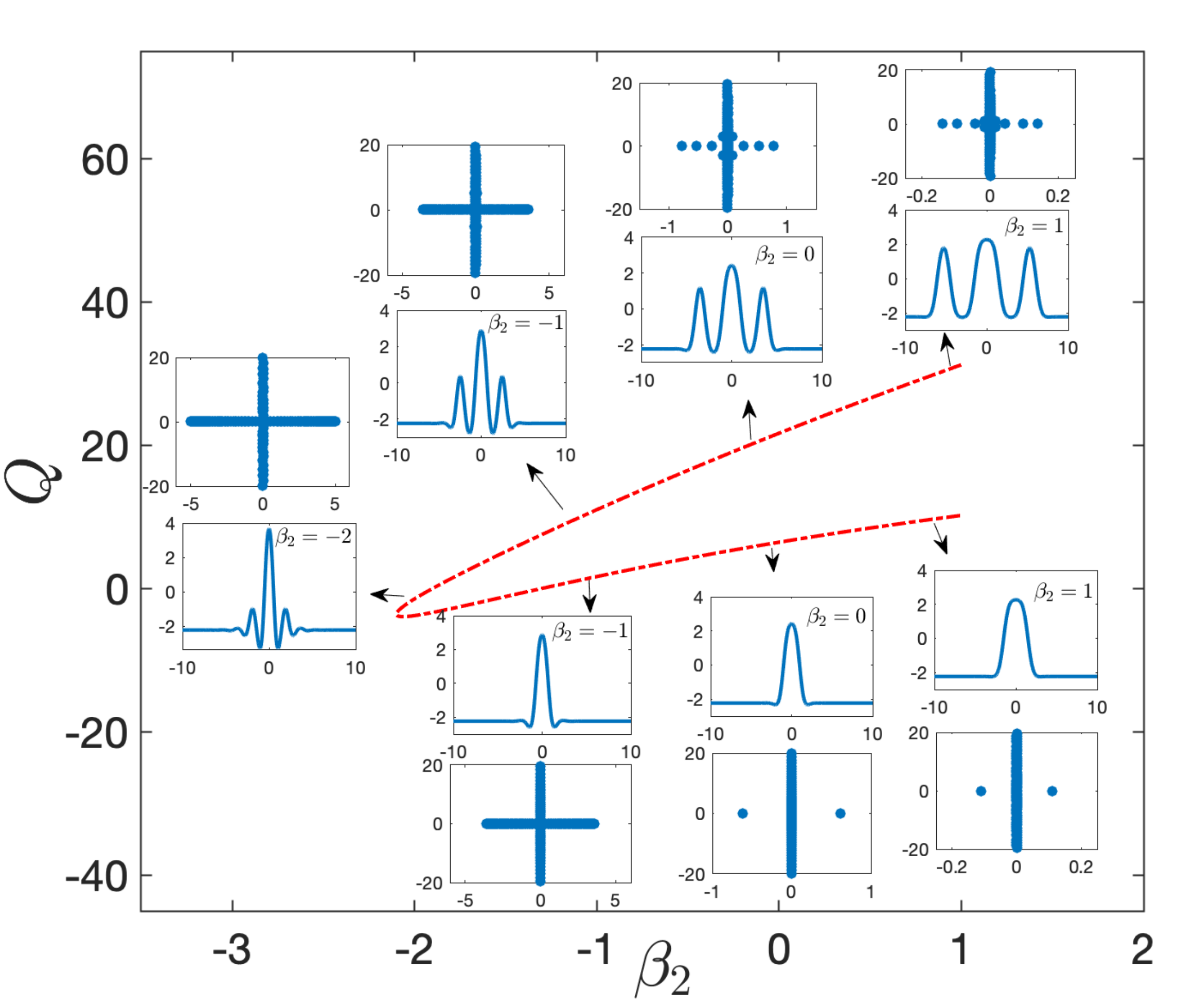}} 
\subfigure[]{
\includegraphics[scale=0.37]{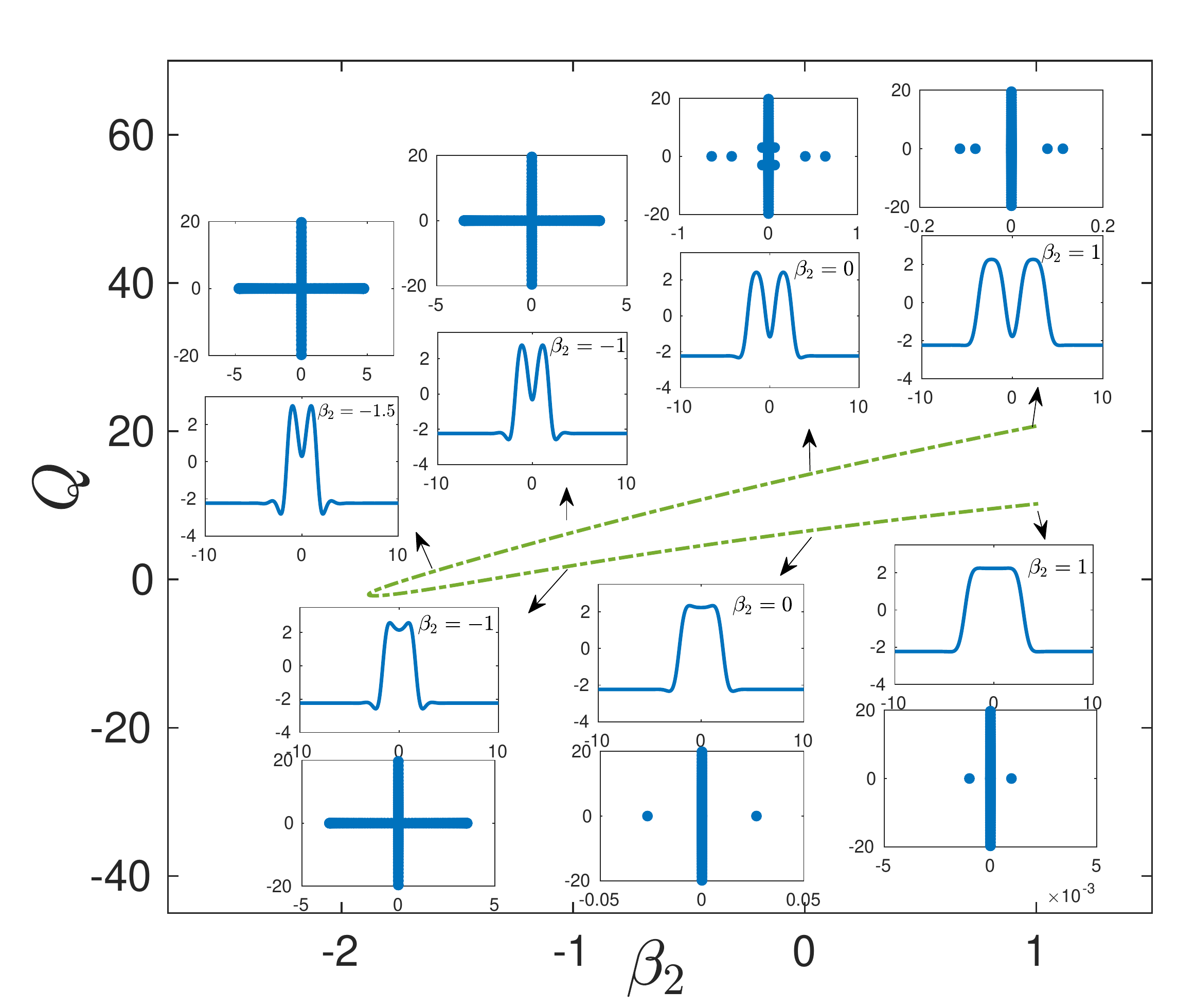}} 
\subfigure[]{
\includegraphics[scale=0.37]{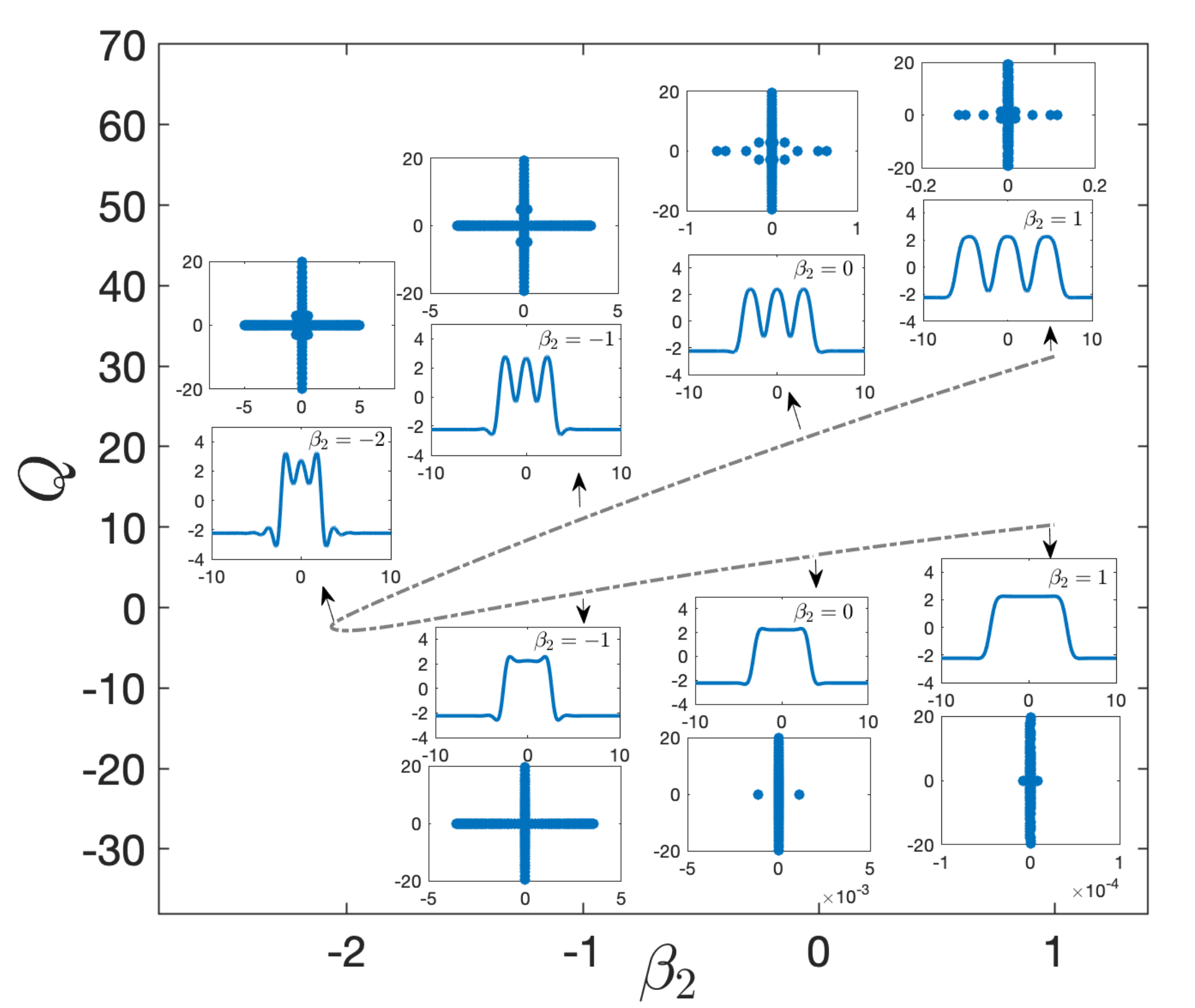}}
\end{center}
\caption{Bifurcation diagram and the corresponding steady state solutions and spectrums for a) family 1, b) family 3, c) family 5. }
\label{fig:bif_plots_13}
\end{figure}

These families are grouped together as they concern
{\it unstable} saddle configurations in their respective
lower branches, as is clearly manifested in each of the bottom
right insets in panels (a)-(c) in Fig.~\ref{fig:bif_plots_13}.
Indeed, in each case the two-kink configurations pertain
to the first, the second and the third local maxima of the
energy landscape associated with the two 2-kink states
which means that their out-of-phase motion should give
rise to a dynamical instability.
Consequently the two bottom right insets in each branch
feature the associated real pair with a corresponding
eigenmode that should dynamically destabilize the relevant
state. As we move towards negative $\beta_2$, once again
the modulational instability discussed previously takes
place and all relevant configurations are unstable due
to the continuous spectrum portion lying along the real axis.

\begin{figure}[tbp]
\begin{center}
\subfigure[]{
\includegraphics[scale=0.24]{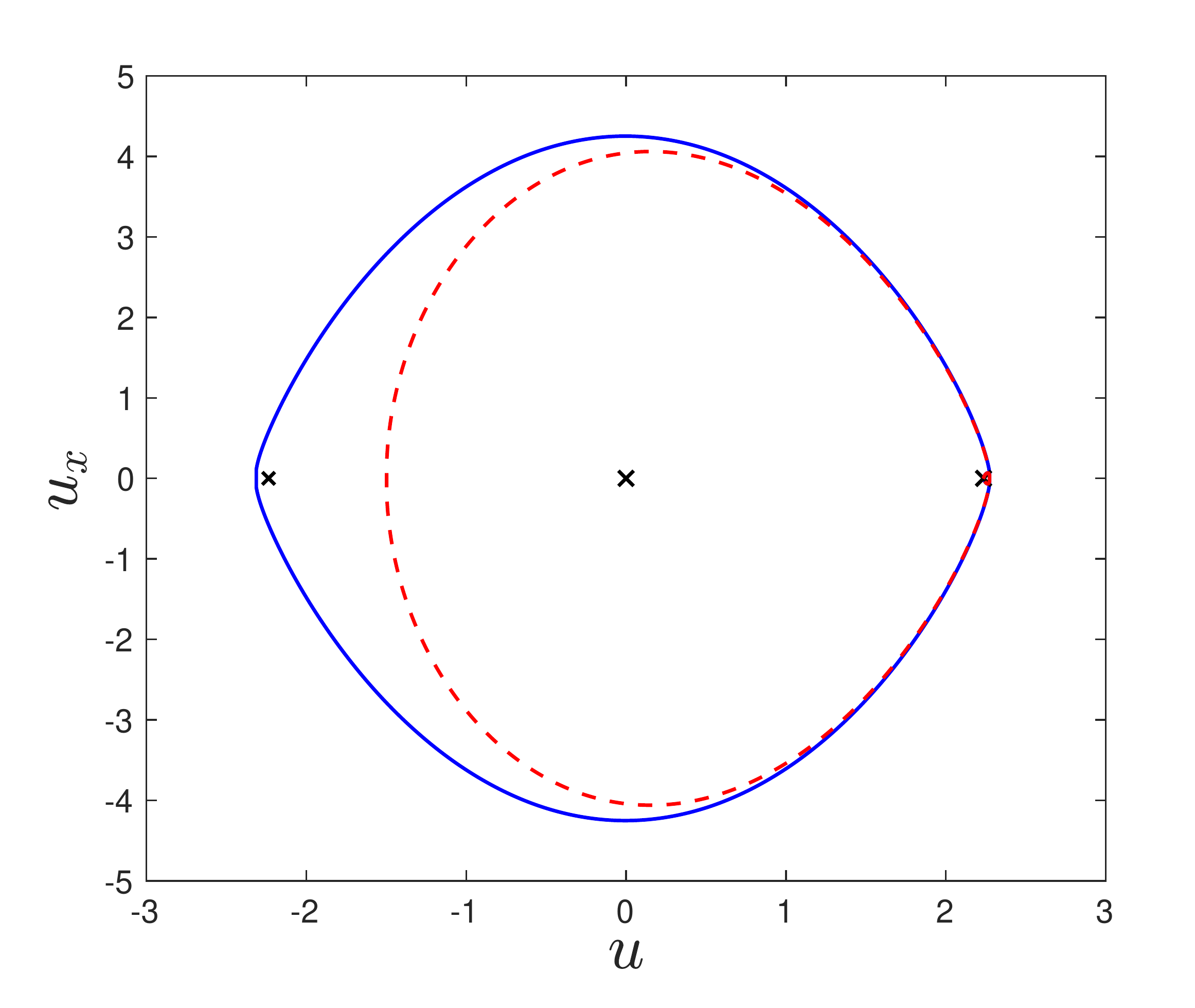}} 
\subfigure[]{
\includegraphics[scale=0.24]{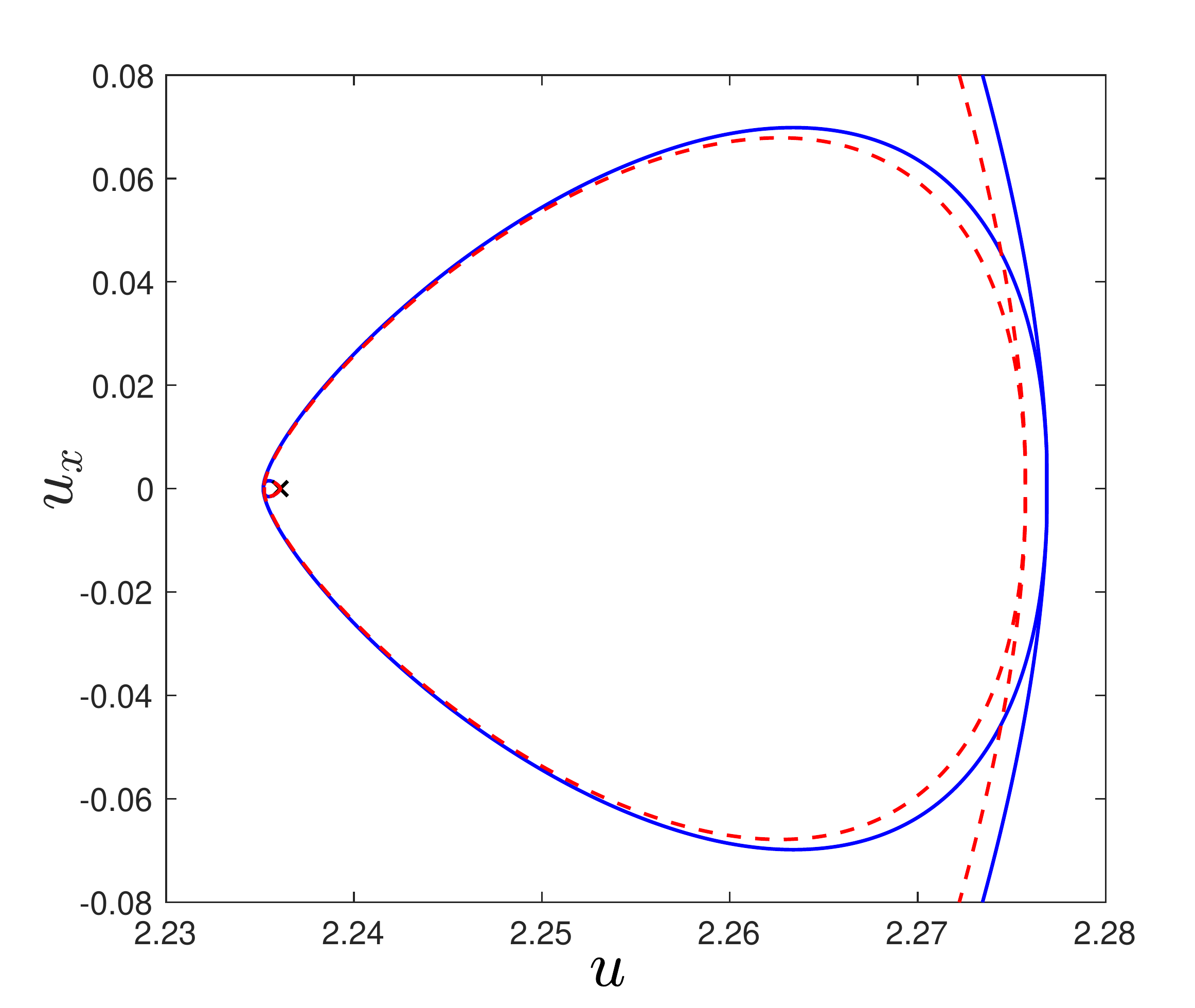}} 
\subfigure[]{
\includegraphics[scale=0.24]{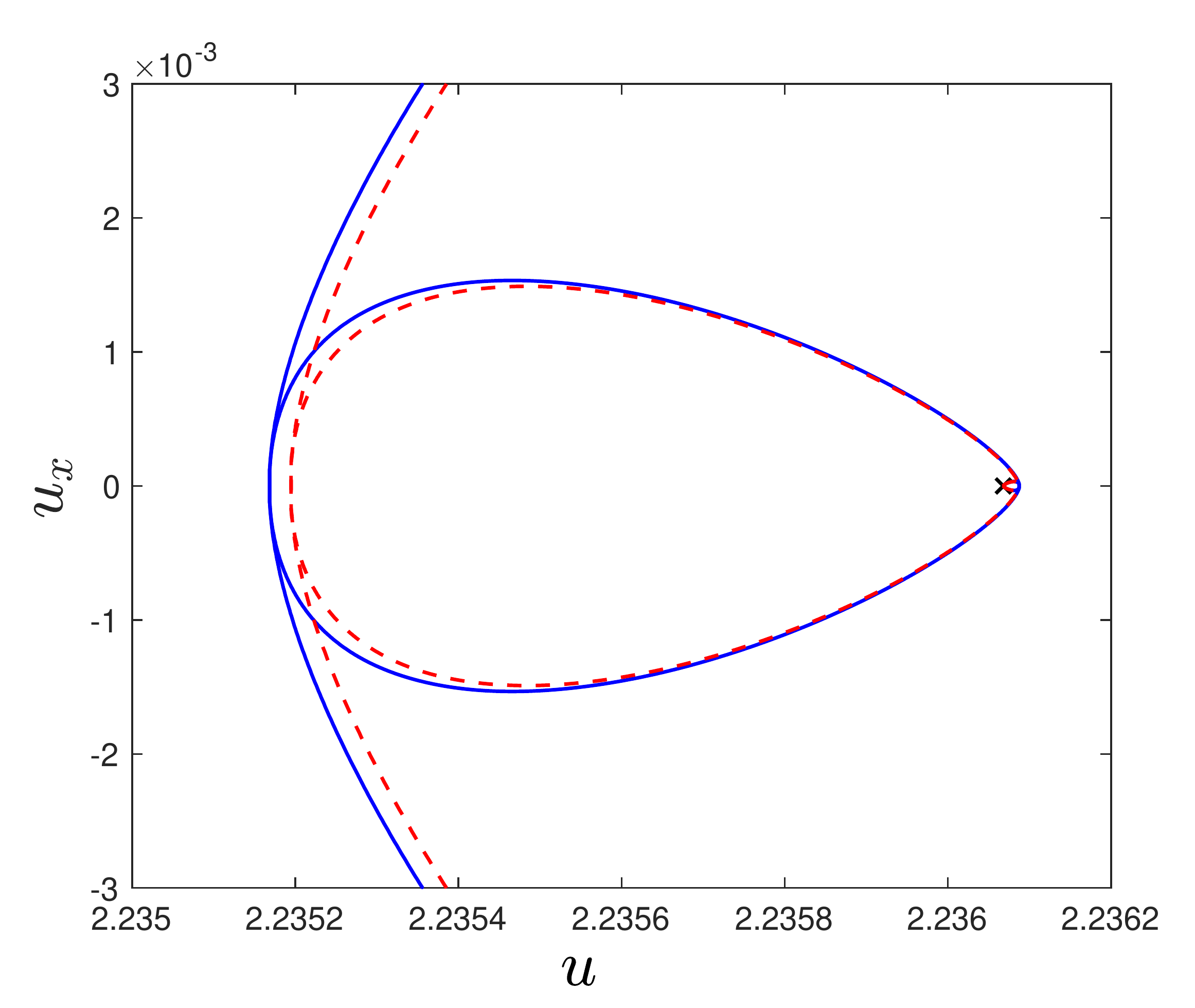}} 
\end{center}
\caption{(a) Phase potraits for family 0 (dashed line) and family 1 (solid
line) in the plane of $u$-$u_x$. Panel (a) shows a larger scale, while
panels (b) and (c) manifest zooms near the right fixed point. One can see the
resulting loops that are associated with the exponentially decaying in amplitude
oscillatory tails connected with the saddle-spiral fixed point.}
\label{fig:phase_plots}
\end{figure}

\begin{figure}[tbp]
\begin{center}
\subfigure[]{
\includegraphics[scale=0.24]{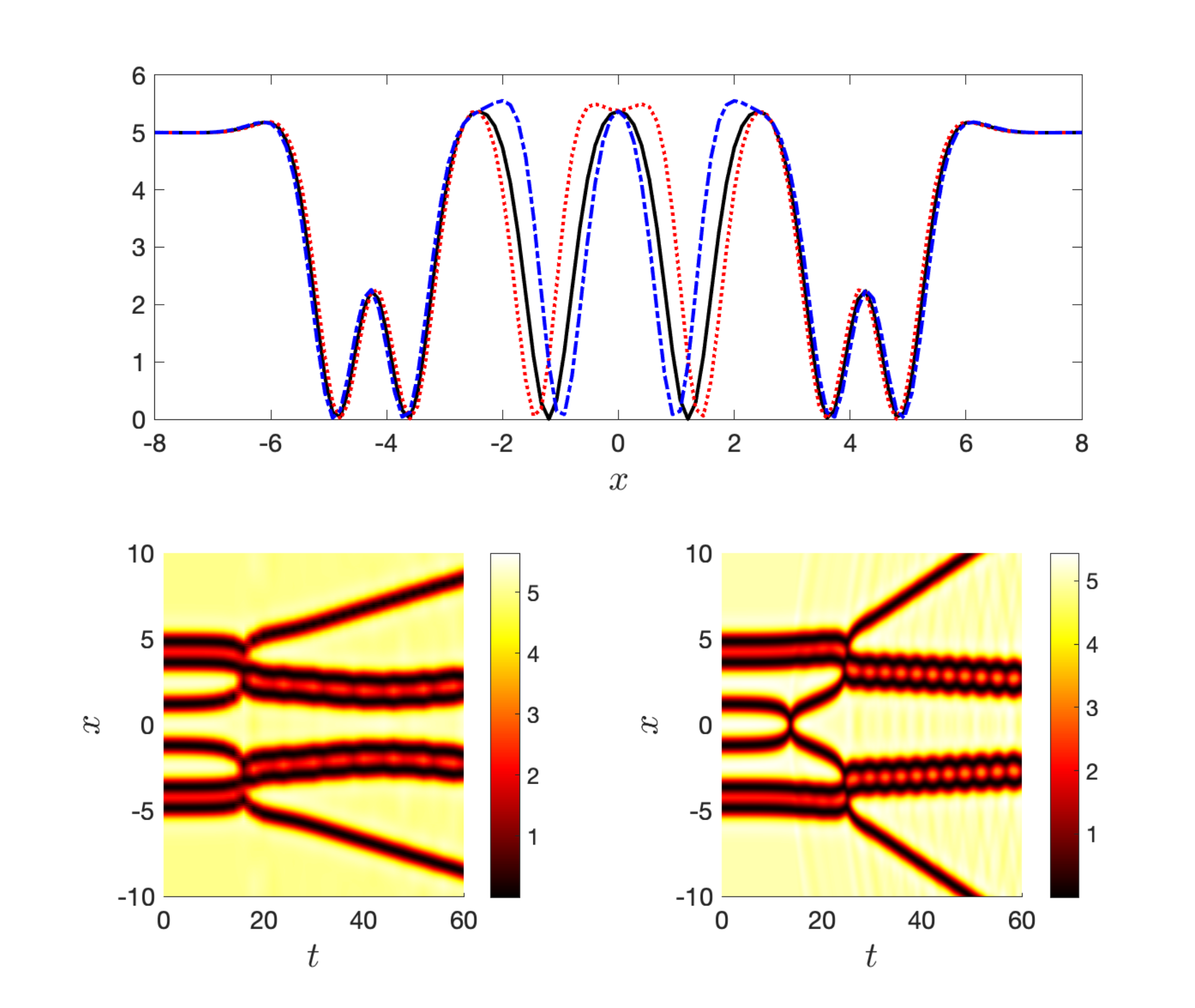}}
\subfigure[  ]{
\includegraphics[scale=0.24]{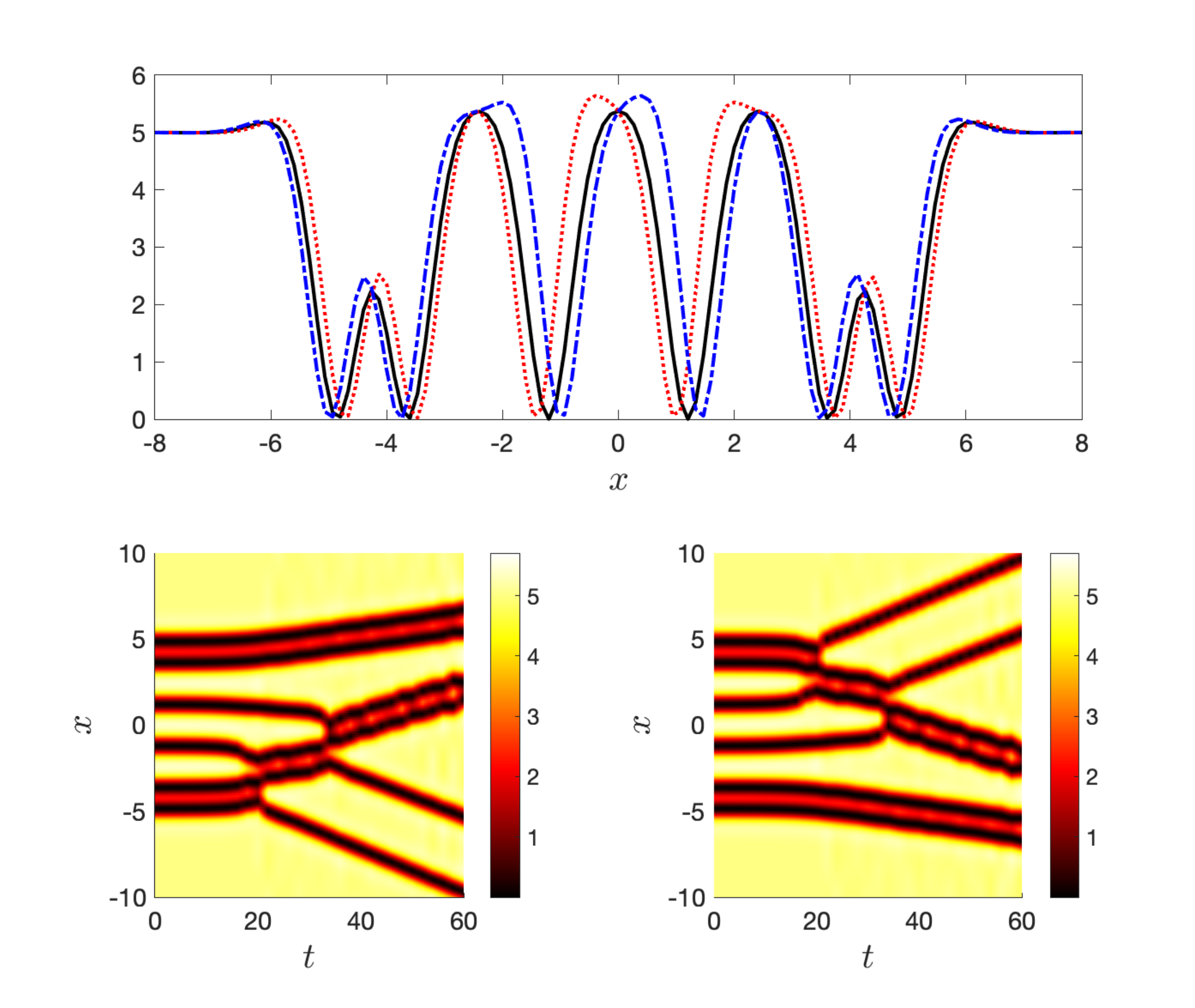}}
\subfigure[  ]{
\includegraphics[scale=0.24]{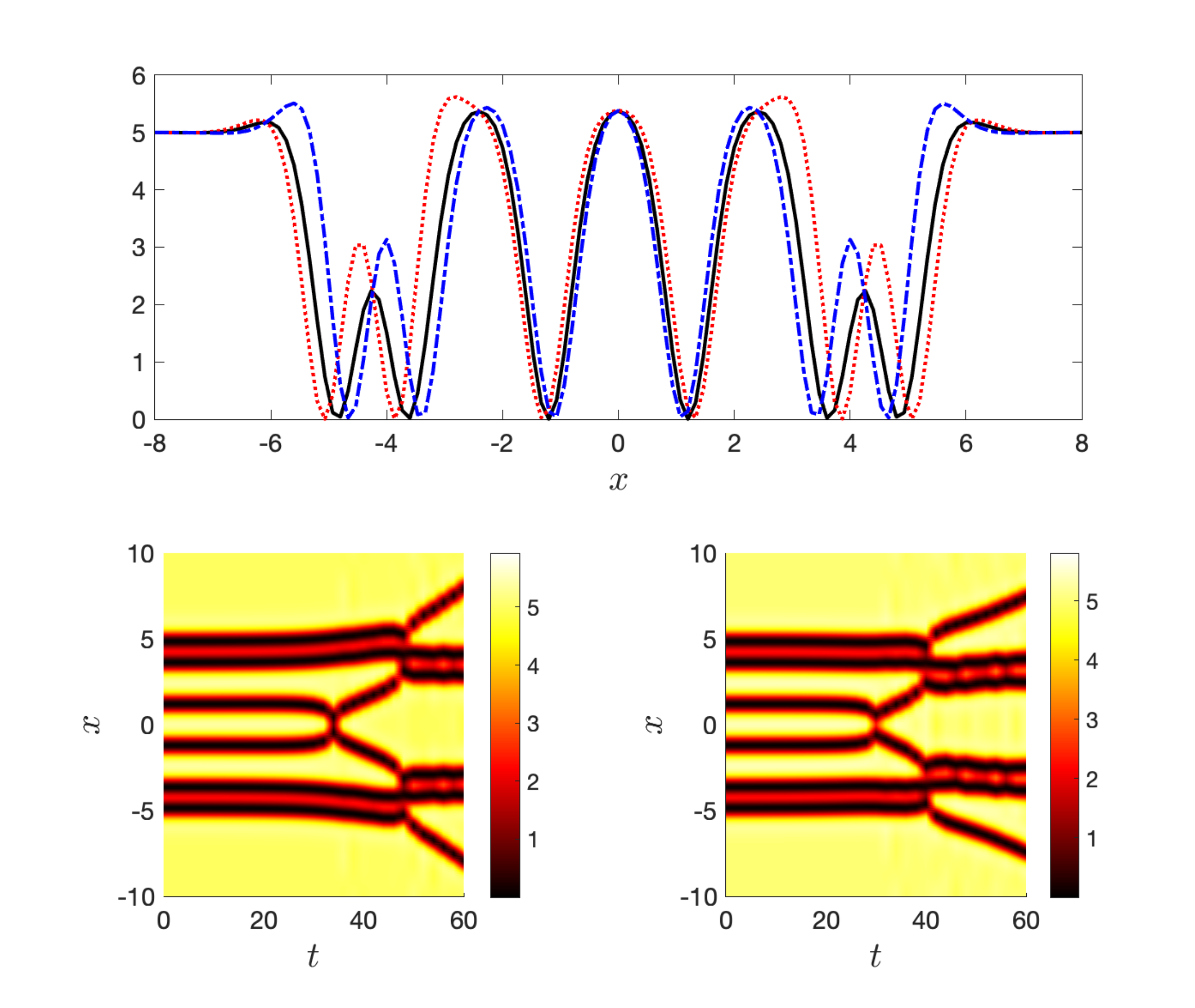}}
\subfigure[  ]{
\includegraphics[scale=0.28]{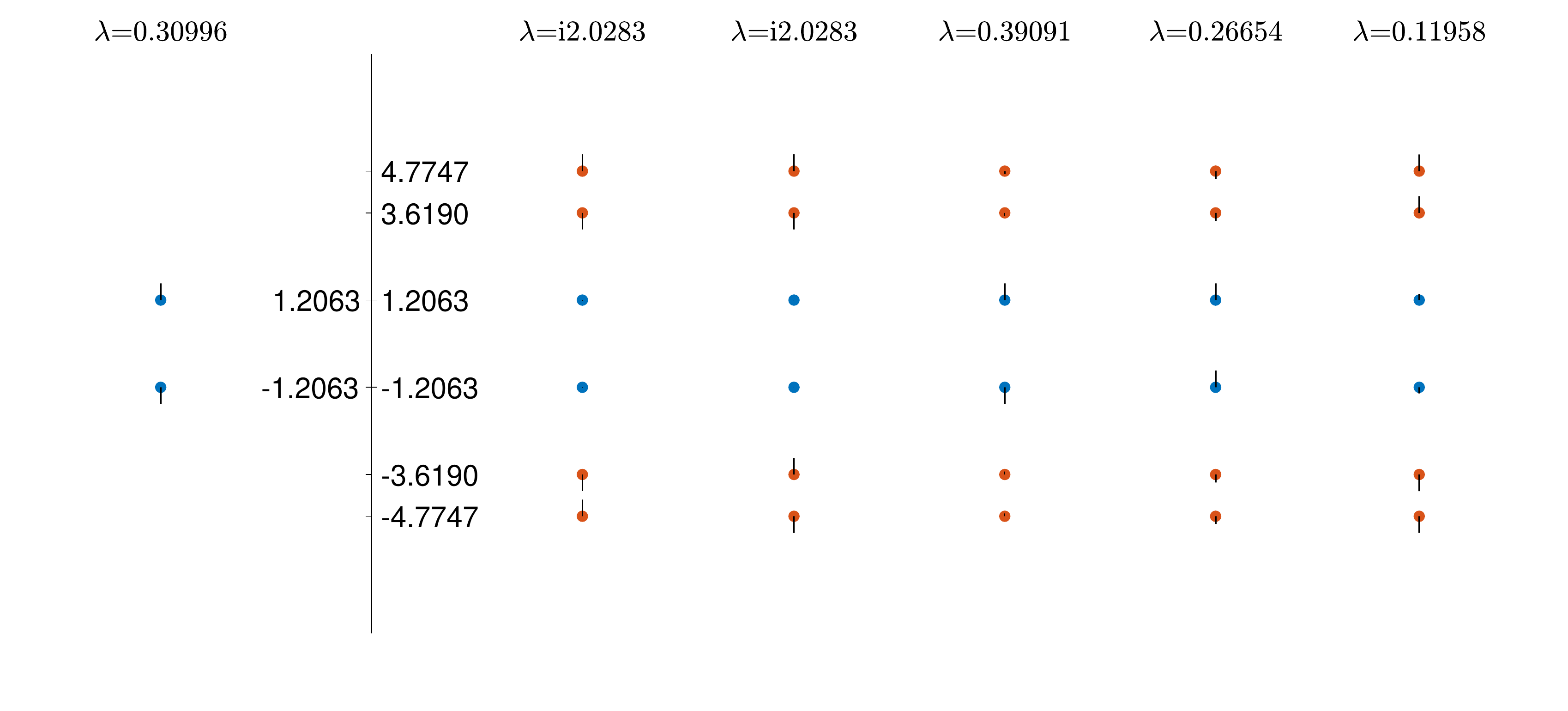}}

\end{center}

\caption{PDE and ODE initial conditions and dynamics for family 1 (upper branch only for PDE). The upper figures of (a), (b), (c) show the plots of $\|u\|^2$ (black), $\|u+v_j\|^2$ (red), and $\|u-v_j\|^2$ (blue) for $\beta_2=0.5$. $v_j$ is the eigenfunction corresponding to the PDE real eigenvalue $\lambda_j$, with (a) $\lambda_1$: 0.3808  (b)  $\lambda_2$: 0.2638 and (c) $\lambda_3$: 0.1263. For each of (a), (b), (c), the lower left figure is the contour plot that results from using $\|u+0.01v_1\|^2$ as the initial condition, and the lower right figure is the contour plot that results from using $\|u-0.01v_1\|^2$ as the initial condition. (d) gives the ODE values for the soliton positions (left of the vertical line is the lower branch, right of the vertical line is the upper branch) and the ODE eigenvalues along the top (again lower branch shown left, and upper branch shown right). Arrows on the points indicate the initial directions of the solitons (all directions would be reversed if $v_1$ is replaced by $-v_1$.)
  }

\label{fig:dyn_fam_1}
\end{figure}

Each of the relevant families features a turning point
beyond which we move to the upper portion of the corresponding
branches. These branches feature 6 kinks for the families 1 and 5
and 4 kinks for the family 3. As discussed previously, family 3,
along with family 0 are special in this regard, while all other
families feature 6 kink states in their upper portions.

{To better understand this we need to turn our focus to the single kink bifurcation diagram. As we move to the left along the lower branch (and \(\beta_2\) decreases), undulations on the oscillating tails on both sides of the kink increase in size. When the largest undulation (of
which there are two, one on each side, due to symmetry) reaches a critical size, we reach the turning point. Then as we move on to the upper branch, (and now \(\beta_2\) increases), the undulations decrease in size, except for the two largest ones. These continue to grow and eventually give rise to the two new kink pairs, one on each side of the kink}.

{Two-kink solutions behave in a similar manner, but the number and the location of the new kink pairs depend on the distance between the two original kinks. That distance can only be such that the oscillating tails between the two kinks interfere either constructively (odd families) or destructively (even families). In the latter case, the two largest undulations will be the ones outside of the kink pair and these will give rise to the two new kink pairs. In the former case, the two largest undulations will be the ones inside the kink pair, {\it if} there is enough space to do so, as in families 5, 7, 9... , where two new pairs appear in the region inside the two kinks. In the case of family 3, there is space for only one new pair to appear, while in the case of family 1 there is no space for any such pair at all, so the new pairs can appear from the undulations outside the two original kinks. }

{The interference between the oscillating tails of the two kinks can also explain the differences between the value of \(\beta_2^{cr}\) of each family. In the constructive case, the largest undulations reach their critical size earlier, as we move to the left. So we reach the turning point for larger \(\beta_2\), compared to the single kink case. In the destructive case, on the other hand, undulations are smaller so we need to move further to the left for them to reach their critical size. So the turning point corresponds to \(\beta_2\) smaller than the one in the single kink case (with the notable exception of family 1). }

It should be added here that for the families 0 and 1, we have also
depicted the phase portrait of the plane $(u,u_x)$ of Fig.~\ref{fig:phase_plots}.
The aim of the figure is to showcase how for family 0, the kink-antikink profile
only loops around $0$, but does not make it to loop around the (spatial) fixed
point of $-\sqrt{\mu/\gamma}$, while in the case of family 1, that looping does
(as the first such example among the families) take place. The effectively self-similar
pattern of the spatial configuration as it loops around the saddle-spiral fixed point
at $u=\sqrt{\mu/\gamma}$ is further illustrated in the zooms of panels (b) and (c).

Turning now to the upper branch of family 1, the analysis of the relevant 
state is conveyed in Fig.~\ref{fig:dyn_fam_1}.
Panel (d) summarizes our theoretical predictions.
More concretely, the 6-kink state features 2 oscillatory
modes and 3 real ones, in addition to the neutral translational one. 
The one with the largest growth rate
($\approx 0.391$) features an out-of-phase motion of the
two innermost kinks, while the other 4 remain essentially
immobile. This instability is showcased in the dynamical evolution of
panel (a) where
we see that these inner kinks may either move outward
colliding with the other two pairs (and forming breathing
pairwise bound states, while the outermost kink is expelled)
or they may move inward, collide and then move outward
again, leading to the same fate as the previous example.
We have also excited the two other unstable modes
in panel (b) (for growth rate $\approx 0.267$) and panel
(c) (with growth rate $\approx 0.12$), respectively.
In the former case, the two inner kinks move in one
direction, while the four outer ones move in the opposite
direction. In both shown examples of panel (b), 
this leads to collisions
and pairwise formations of one kink with a bound state
pair. In each example where this happens, there is a 
``change of allegiance''. The pair member closest to the
single kink now forms a bound state with the formerly
single kink, while the pair member furthest from the single
kink is now ``freed'' and moves in the direction that the single
kink used to move. In the case of the eigenmode excited in panel (c)
the outer kinks move outward or inward, while
the centermost pair stays put. However, what ends up being observed
is more akin to the dynamics of panel (a), which appears to be the dominant instability, since the associated eigenmode growth rate is a factor of (nearly) 4 times larger than the eigenmode initially excited in panel (c).  See also the discussion surrounding the case example shown later in Fig.~\ref{fig:projections}(e).

\begin{figure}[tbp]
\begin{center}
\subfigure[  ]{
\includegraphics[scale=0.24]{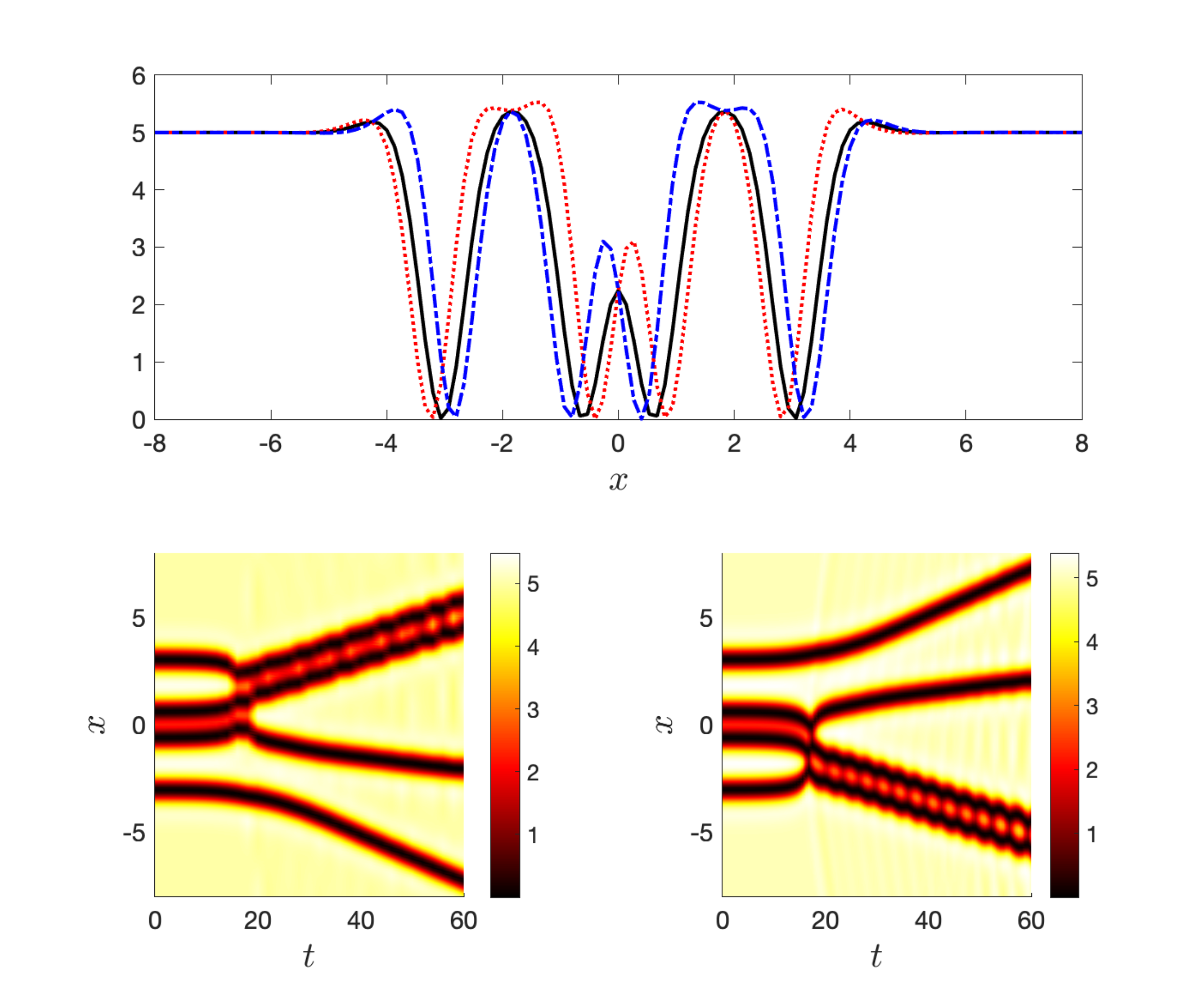}}
\subfigure[  ]{
\includegraphics[scale=0.24]{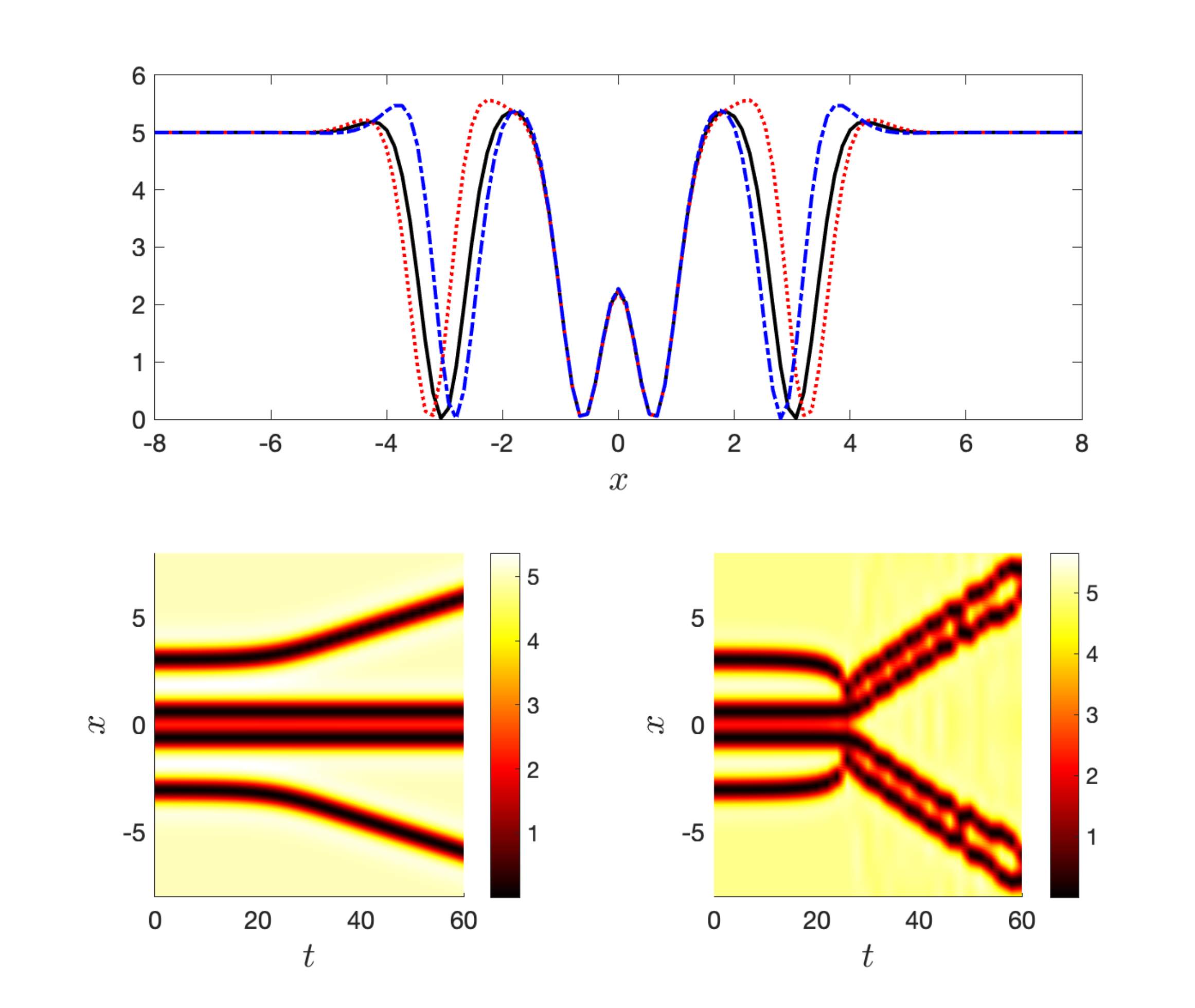}}
\subfigure[  ]{
\includegraphics[scale=0.28]{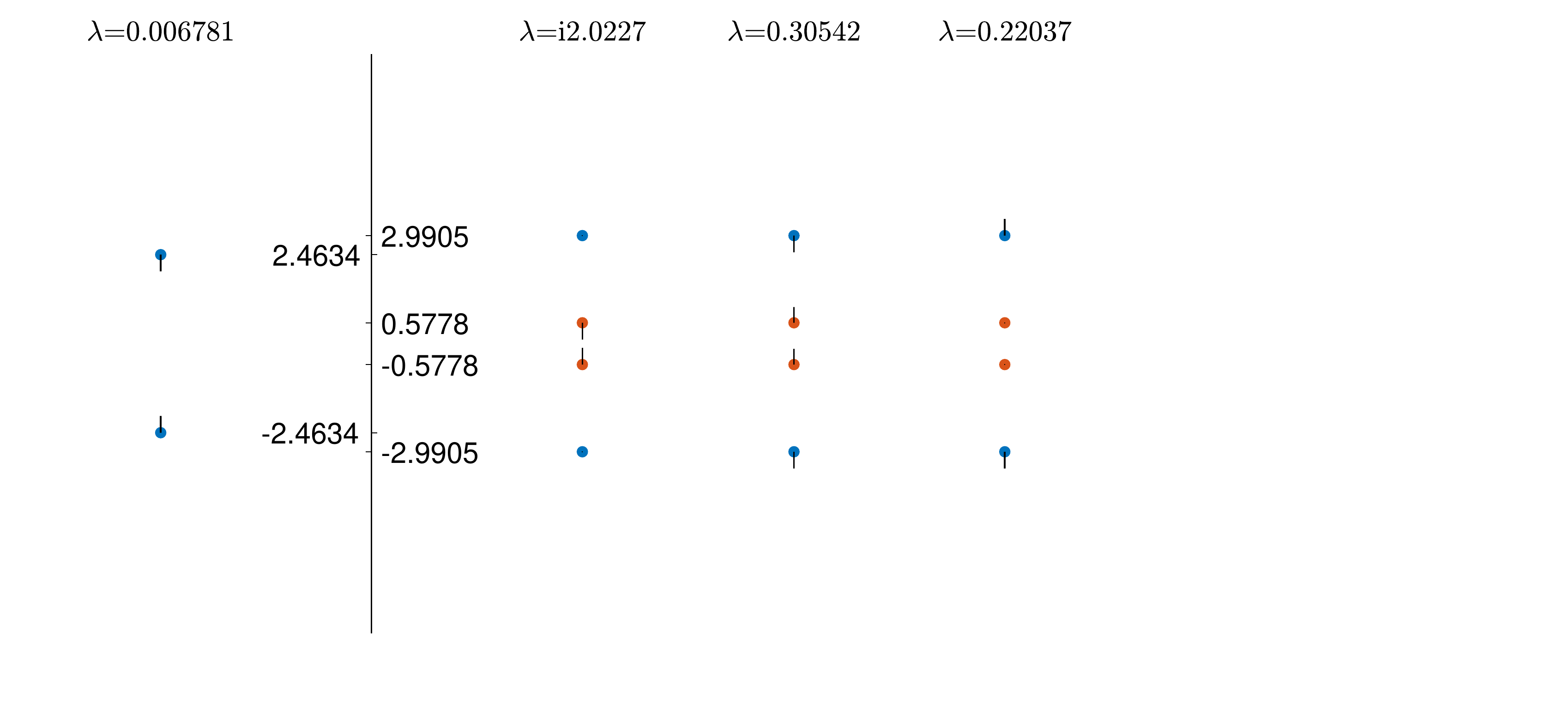}}
\end{center}

\caption{PDE and ODE initial conditions and dynamics for Family 3 (upper branch only for PDE). Similar to Figure \ref{fig:dyn_fam_1}, except for Family 3 instead of Family 1, with $\beta_2=0.5$ and PDE eigenvalues (a) $\lambda_1$: 0.3083  and (b)  $\lambda_2$: 0.2104.
  }
  
\label{fig:dyn_fam_3}
\end{figure}

In the case of family 3, the corresponding
dynamical picture is provided in Fig.~\ref{fig:dyn_fam_3}.
The lower branch situation is again simple (with the
out-of-phase motion of the two kinks predicted in panel
(c) being responsible for the instability of this 
saddle-point configuration at a larger distance
of $\approx 2.46$ at equilibrium. However, as indicated
above, this is an example whereby the upper branch involves
only 4 kinks. In this setting, our effective particle
theory predicts the existence of 2 unstable
modes with growth rates $\approx 0.305$ and
$\approx 0.22$. The largest growth rate involves
the inner kinks moving in one direction and the outer
ones in the opposite, as shown in panel (a) of the
figure. This leads, in line with what we saw 
before, to the collision of the inner pair
with one of the outer kinks, and once again
the same phenomenon of ``change of allegiance''
as discussed above. The less rapid growth
is associated with a mode whereby the inner
kinks stay put while the outer ones move
either outward or inward (depending on the
sign of the perturbation), as shown in panel (b).
Among these cases, the outward motion is more
``benign'' as the outer kinks depart to (in principle)
infinity, while the inner kinks remarkably are
sitting at the equilibrium distance of the lower
branch of family 0 and, hence, will stay at that distance
indefinitely given the stability of the latter configuration.
A far more elaborate scenario takes place
when the outer kinks first move inward.
 In this case they collide with the central kinks
leading to an expulsive event where, pairwise, two sets
of kinks (the upper and lower ones, so to speak)
are expelled outward in a breathing, propagating
state. While this seems like a nearly bound state, the
distance between the kinks appears to be increasing
as they move suggesting that it does not pertain to
a stable configuration. Nevertheless, an exploration
of such breathing, propagating states could be
an interesting topic for future study, as it is outside
the scope of the present work.

\begin{figure}[tbp]
\begin{center}
\subfigure[  ]{
\includegraphics[scale=0.24]{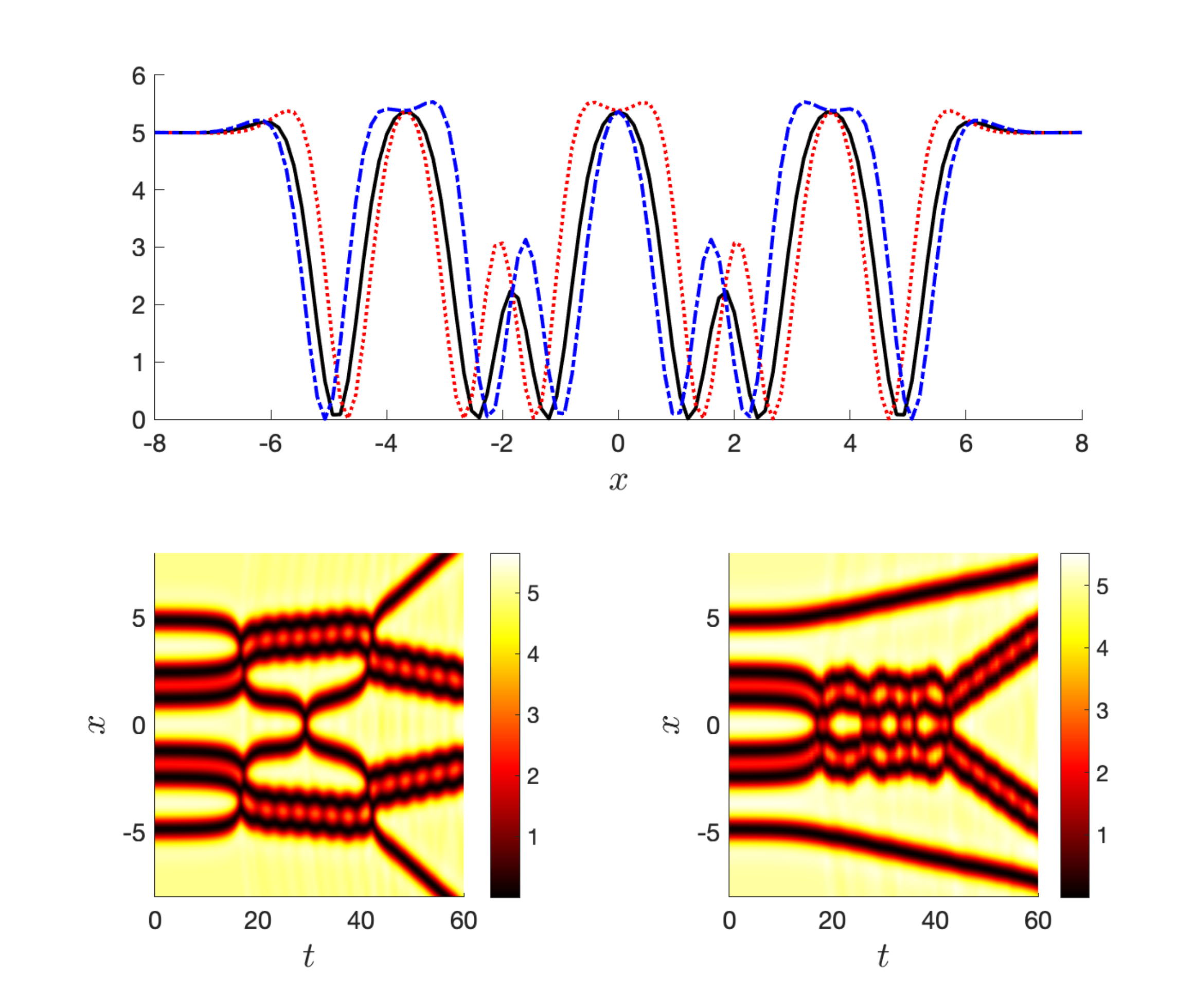}}
\subfigure[  ]{
\includegraphics[scale=0.24]{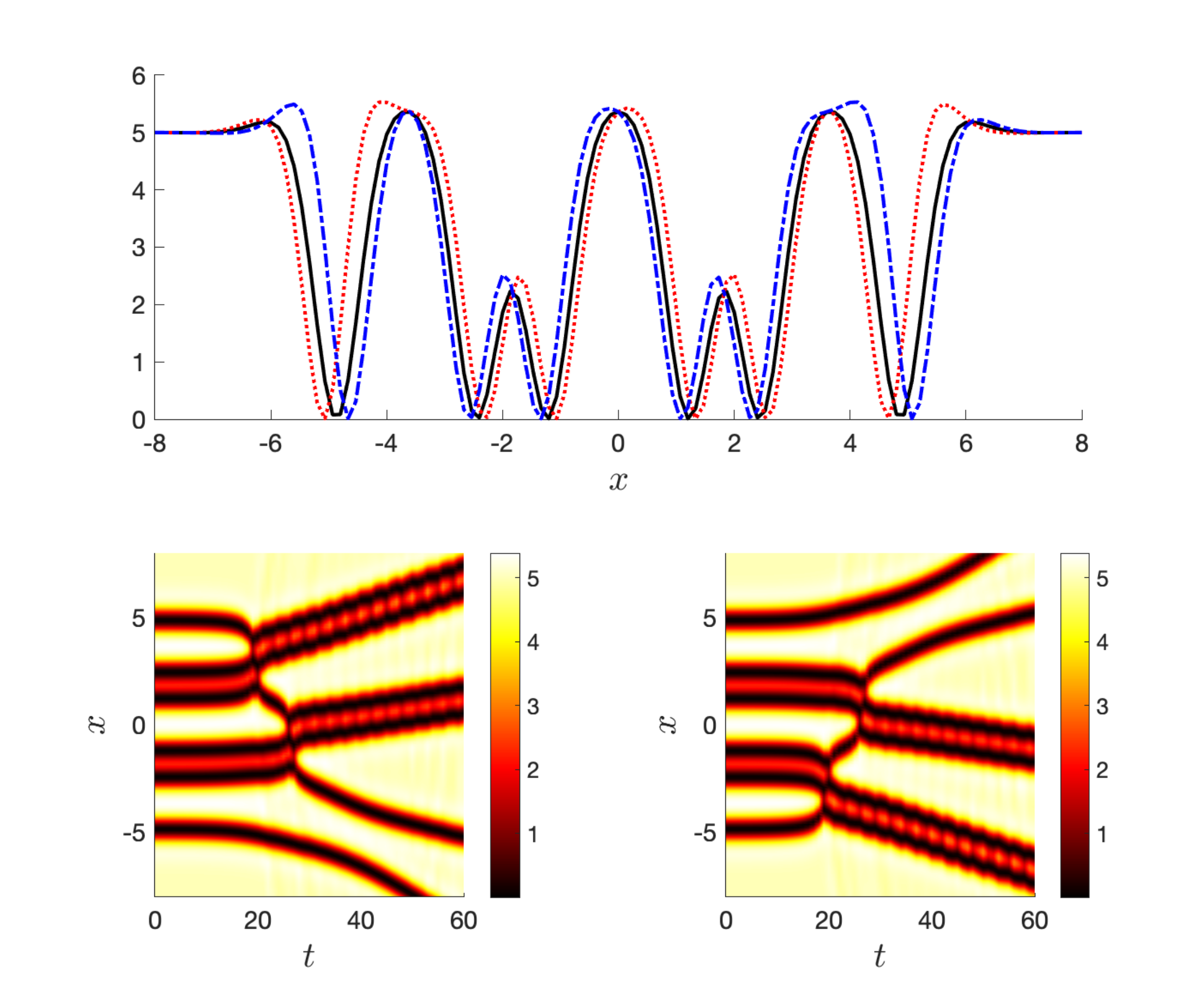}}
\subfigure[  ]{
\includegraphics[scale=0.24]{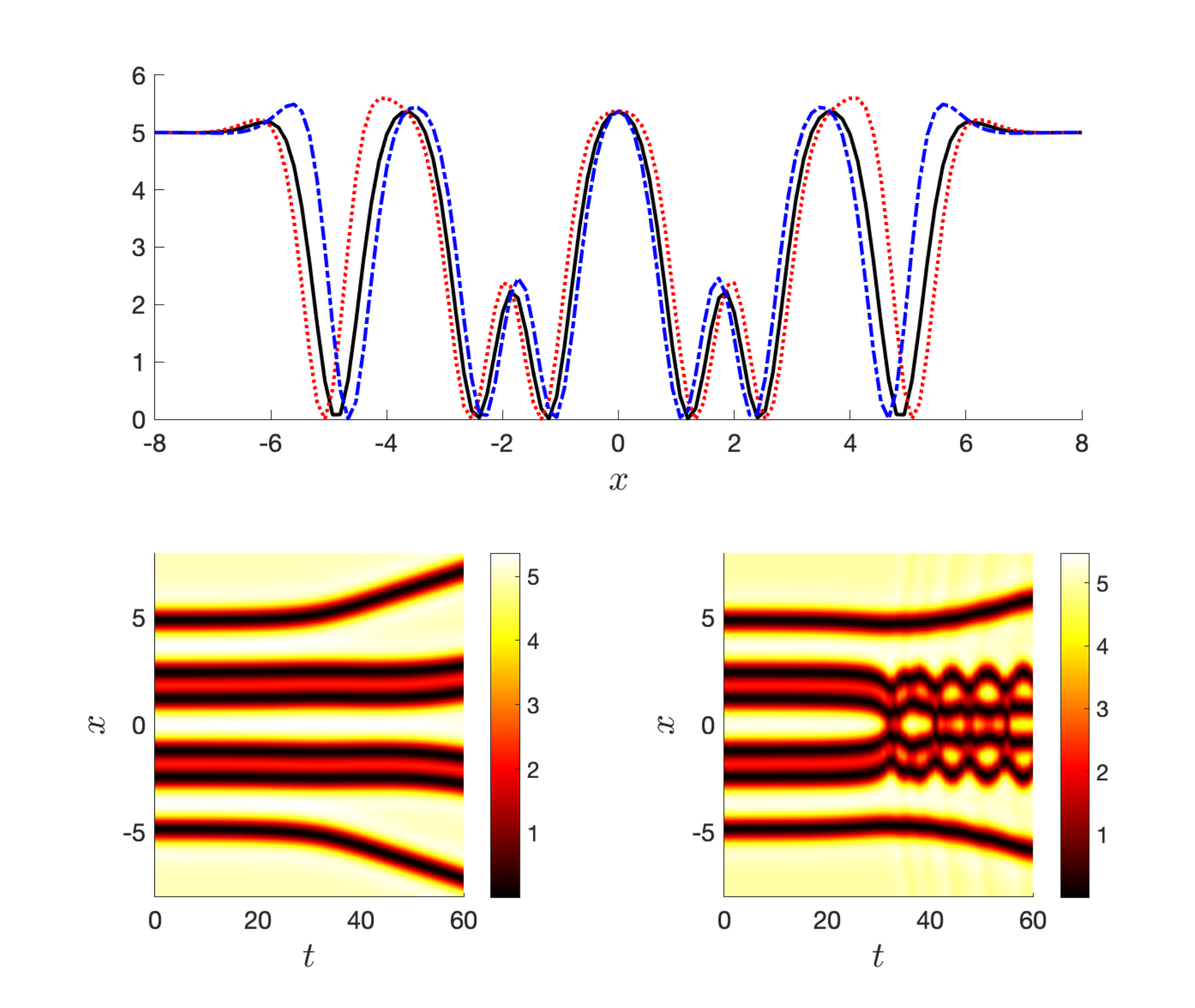}}
\subfigure[  ]{
\includegraphics[scale=0.28]{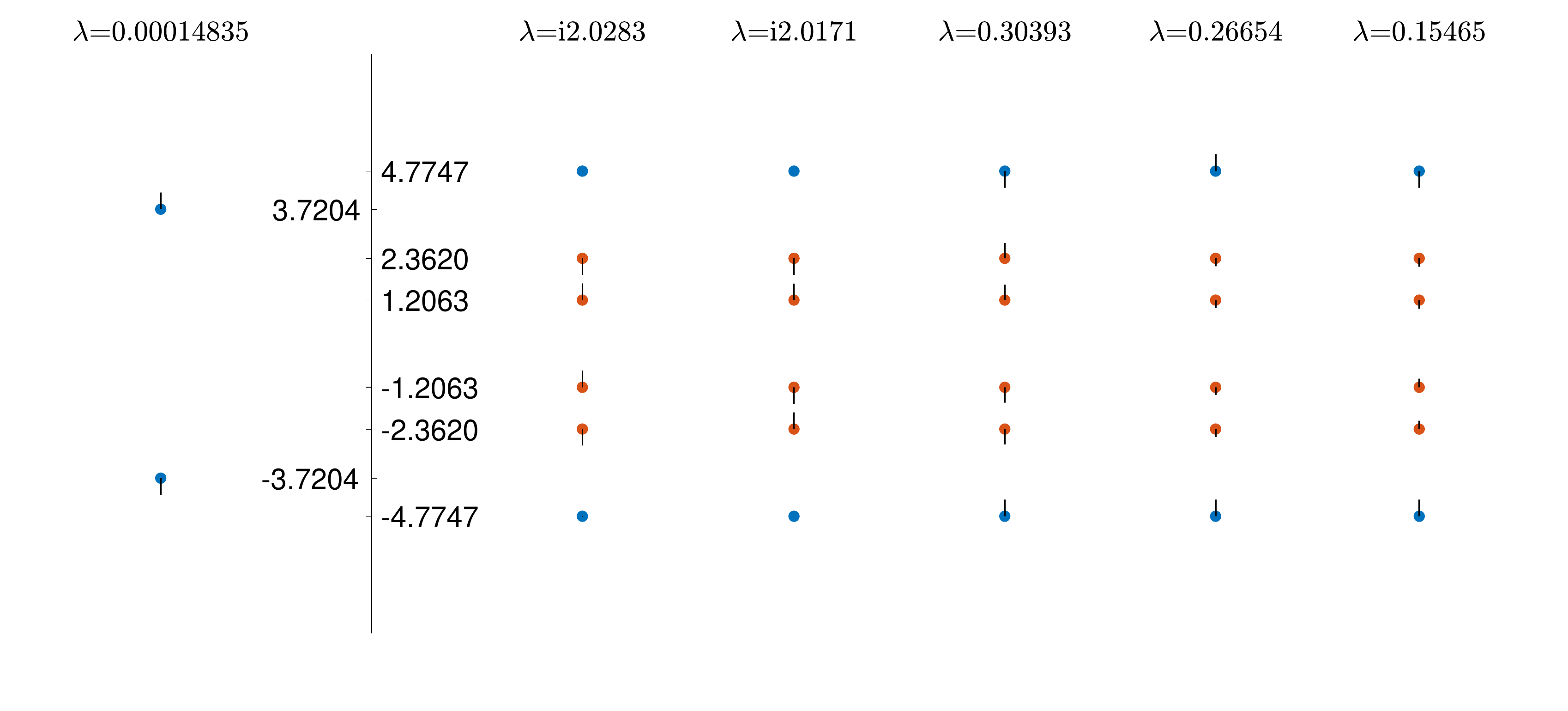}}
\end{center}
  
\caption{PDE and ODE initial conditions and dynamics for Family 5 (upper branch only for PDE). Similar to Figure \ref{fig:dyn_fam_1}, except for Family 5 instead of Family 1, with $\beta_2=0.5$ and PDE eigenvalues (a) $\lambda_1$: 0.3105, (b)  $\lambda_2$: 0.2640, and (c) $\lambda_3$: 0.1509.
  }  
  
   \label{fig:dyn_fam_5}
\end{figure}

We now turn to family 5, which again like most families
has 6 kinks in its upper portion (in addition to 2 
substantially separated kinks at distance of $\approx 3.72$
in its saddle-configuration lower portion). Here again,
we encounter a situation involving 3 unstable modes of
the upper branch, along with 2 oscillatory ones which have
also been included for completeness in Fig.~\ref{fig:dyn_fam_5}, in addition to
the neutral translational mode.
The unstable modes have growth rates of
$\approx 0.304$, $0.267$ and $0.155$ as indicated in
panel (d) of the figure. The most unstable among
these modes involves the out-of-phase motion of
the two inner kink pairs of this configuration and
the opposite to them, also out-of-phase motion of the outer 
kink pair. This can lead, as shown in panel (a) of the figure,
e.g., to a collision of the kink pairs with the outer kinks,
leading to a change of allegiance and then complex
dynamics since the split innermost kinks collide between them
and then again with the breathing pairs (leading to further
change of allegiance etc.). In the case of the two pairs
moving inward they collide with each other, while the
outer kinks move outward. In this case, the complex dynamics
of the 4-kink collision near the center eventually leads,
upon breathing, to two outer moving and breathing
pairs, once again reminiscent of the ones we saw in family
3. Again, such dynamics as well as similar pair breathing
and propagating, for instance,
in panel (b) of the figure are motivating
towards further study of such states. 
The case of panel (b) involves a weak in-phase motion of
the 4 inner kinks and a stronger opposite direction motion
of the 2 outer ones. This leads to a collision of one 
of the outer kinks with one of the inner pairs, and then
a resulting cascade of two changes of allegiance 
as observed in both instances of panel (b) resulting eventually
in two breathing pairs moving in one direction and two
isolated kinks in the opposite direction. 
Finally, the weakest unstable mode of panel
(c) involves all kinks for $x>0$ moving in the same
direction and similarly all those for $x<0$ moving in the 
opposite direction. In the first example of panel (c)
this leads to no collisions with the kinks continuing to
move in their original direction. In the second example,
all kinks move towards the center and the two pairs
collide there, leading to a breathing long-lived excitation,
while the outer kinks initially moving inward are eventually
led, through interaction (and perhaps the dominance of the most unstable
mode of the highest growth rate) to move in the opposite direction,
diverging away from the center.

\begin{figure}[tbp]
\begin{center}
\subfigure[  ]{
\includegraphics[scale=0.24]{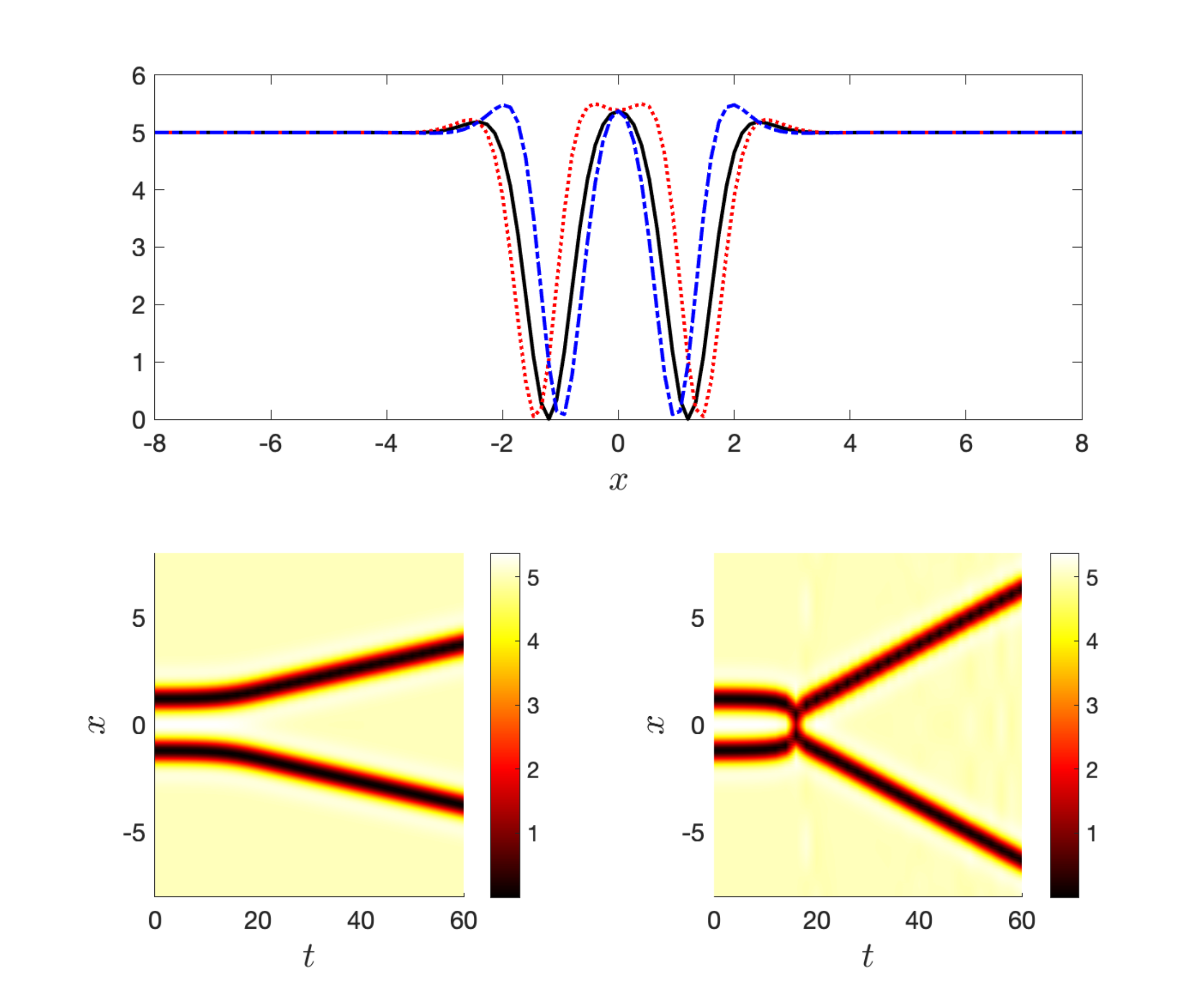}}
\subfigure[]{
\includegraphics[scale=0.24]{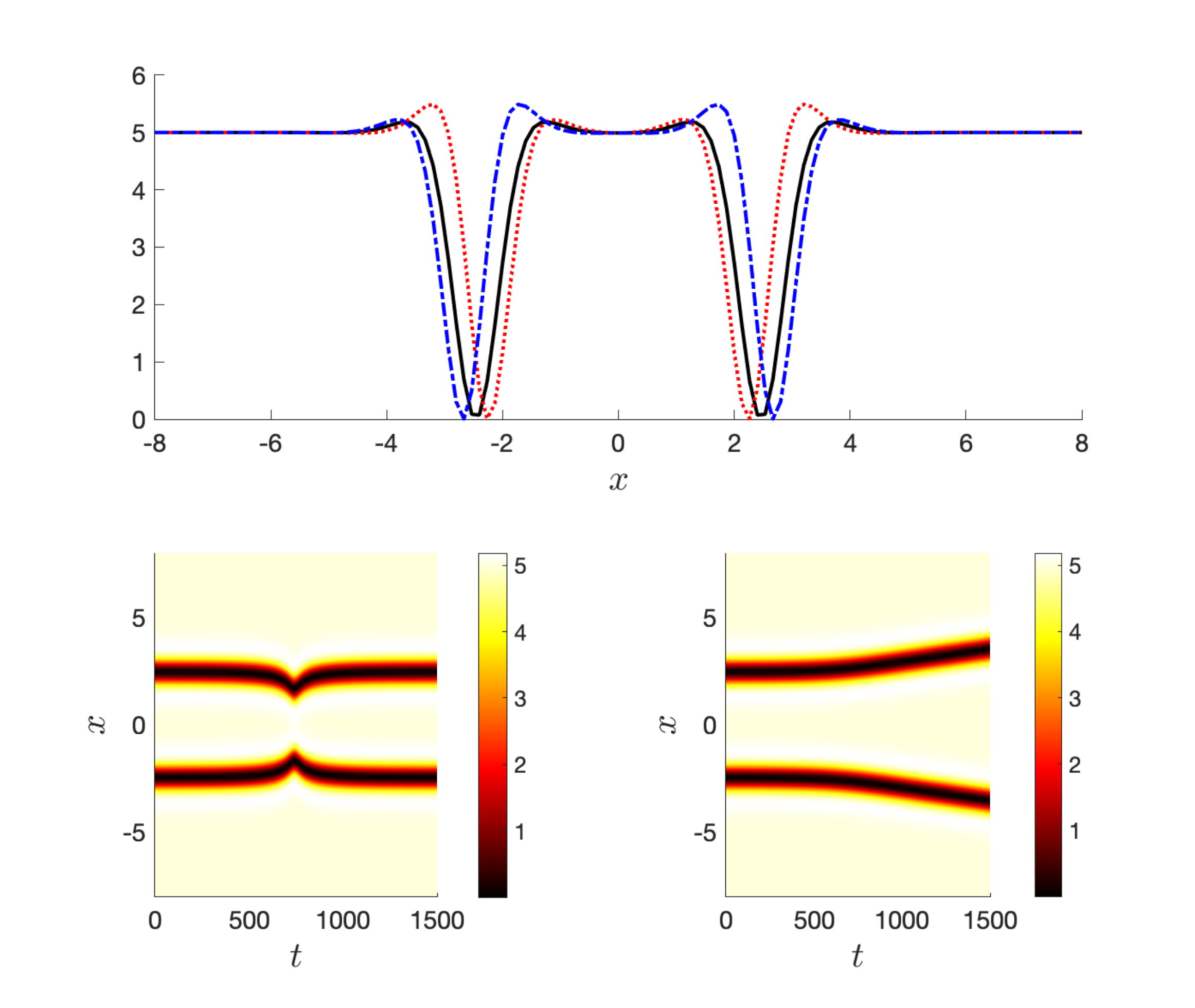}}
\subfigure[]{
\includegraphics[scale=0.24]{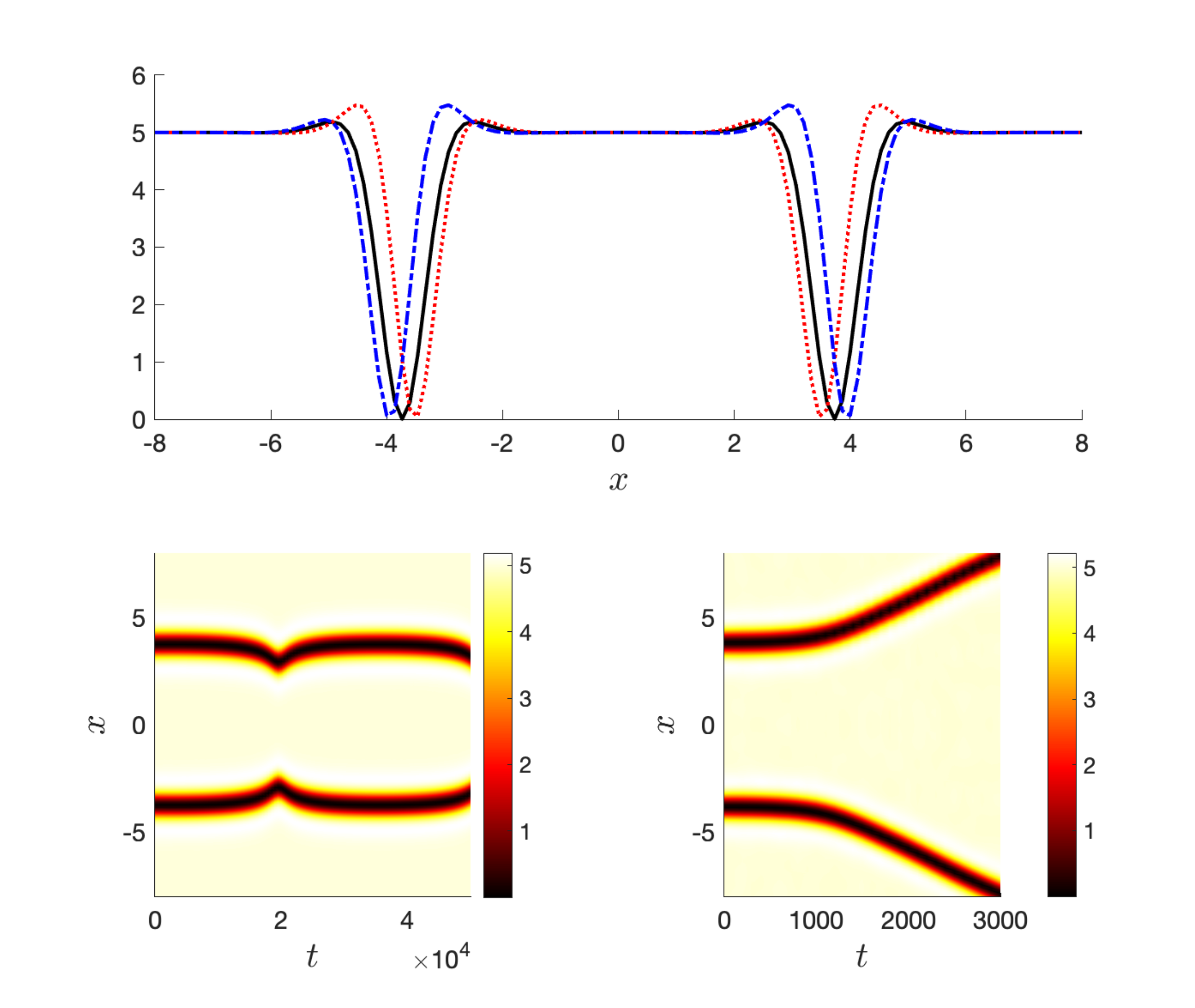}}
\end{center}

\caption{PDE initial conditions and dynamics for lower branches of Families 1, 3, 5. Similar to Figure \ref{fig:dyn_fam_1}, except for lower branches instead of upper branches with (a) Family 1, $\lambda_1$= 0.3012, (b) Family 3, $\lambda_1$= 0.0066, (c) Family 5, $\lambda_1$= 0.00014. The only exception is the bottom right panel of (c). 
For the corresponding ODE initial conditions and dynamics see the bottom panels (left of the vertical line) in each of Figures \ref{fig:dyn_fam_1}, \ref{fig:dyn_fam_3}, and \ref{fig:dyn_fam_5}.  }

\label{fig:bif_plots_13_lower}
\end{figure}

We now turn to a description of the lower branches and
their dynamics for these families for reasons of
completeness. This is shown in Fig.~\ref{fig:bif_plots_13_lower}.
Panel (a) in the figure shows the 2 kink dynamics in family
1. The relevant unstable (saddle) configuration either
destabilizes with the kinks moving outward, or does
so with them moving inward (toward the center) colliding
at $x=0$ and then subsequently moving outward.
Similar examples are shown in panel (b) for the case
of family 3. The only difference in this case
is that when the kinks move inward, they encounter
a higher barrier (that imposed by the solution of family 1)
and hence get trapped in the well between family 3 and
family 1. This is the well involving the stable 
solution of the family 2 around which the dynamics
ends up orbiting in the bottom right panel (b).
Finally, a similar phenomenology is present in the
case of panel (c).
Interestingly, in this case, the oscillation is between
family 5 and family 3 unstable saddle configurations, 
which means that the dynamics is orbiting around the center
(stable) configuration of family 4.

\subsection{Families 2 and 4}

\begin{figure}[tbp]
\begin{center}
\subfigure[]{
\includegraphics[scale=0.37]{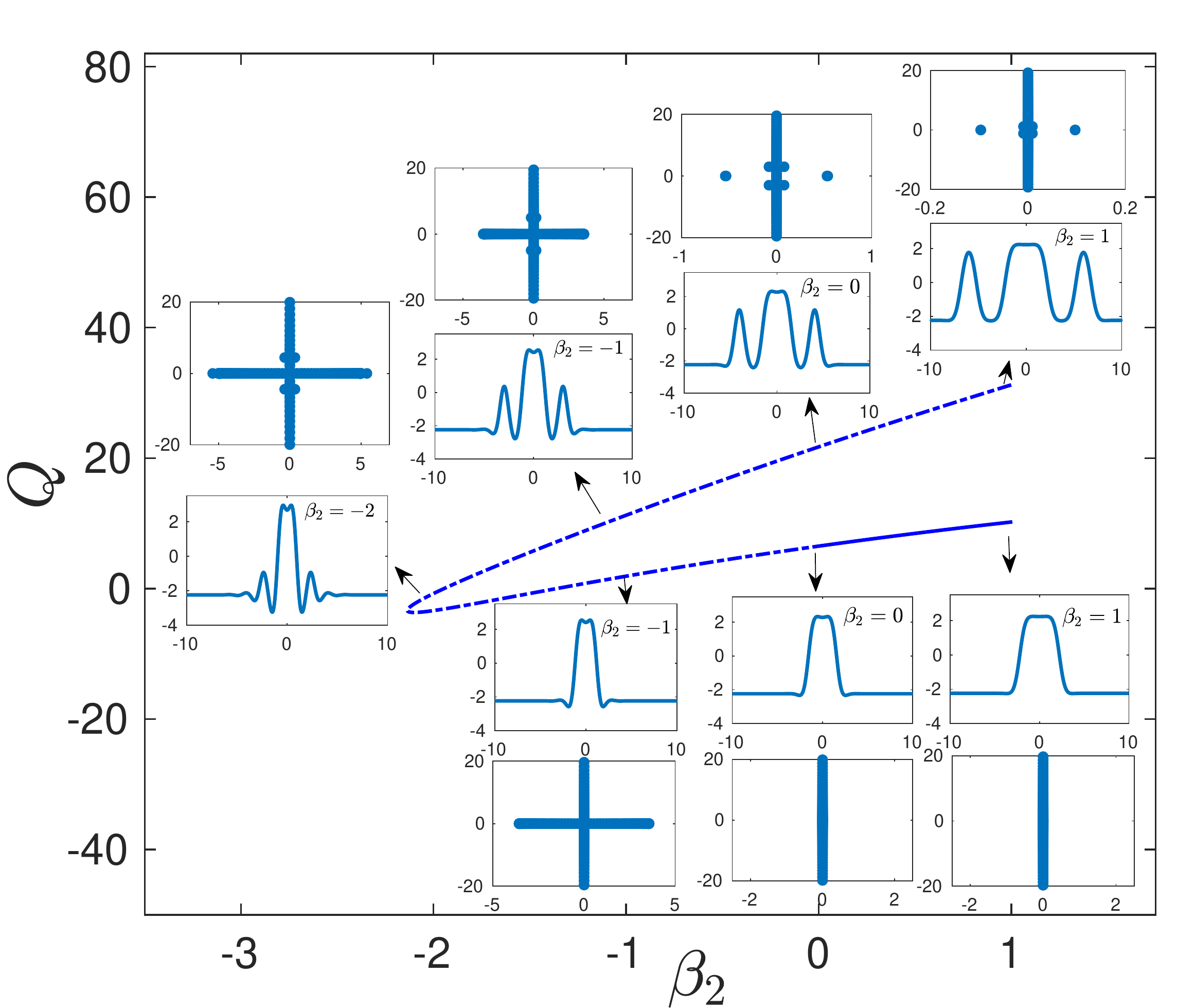}} 
\subfigure[]{
\includegraphics[scale=0.37]{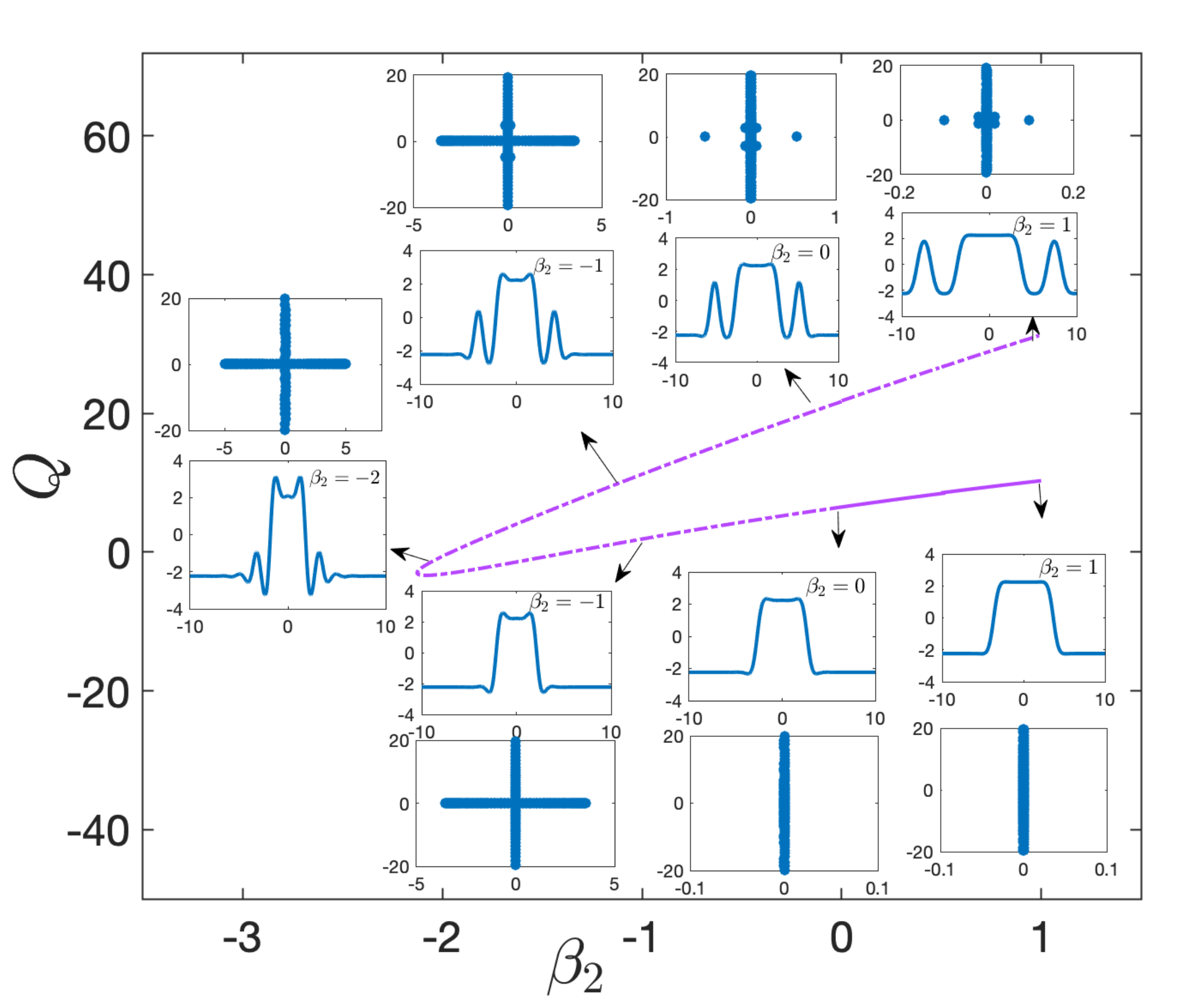}}
\subfigure[  ]{
\includegraphics[scale=0.24]{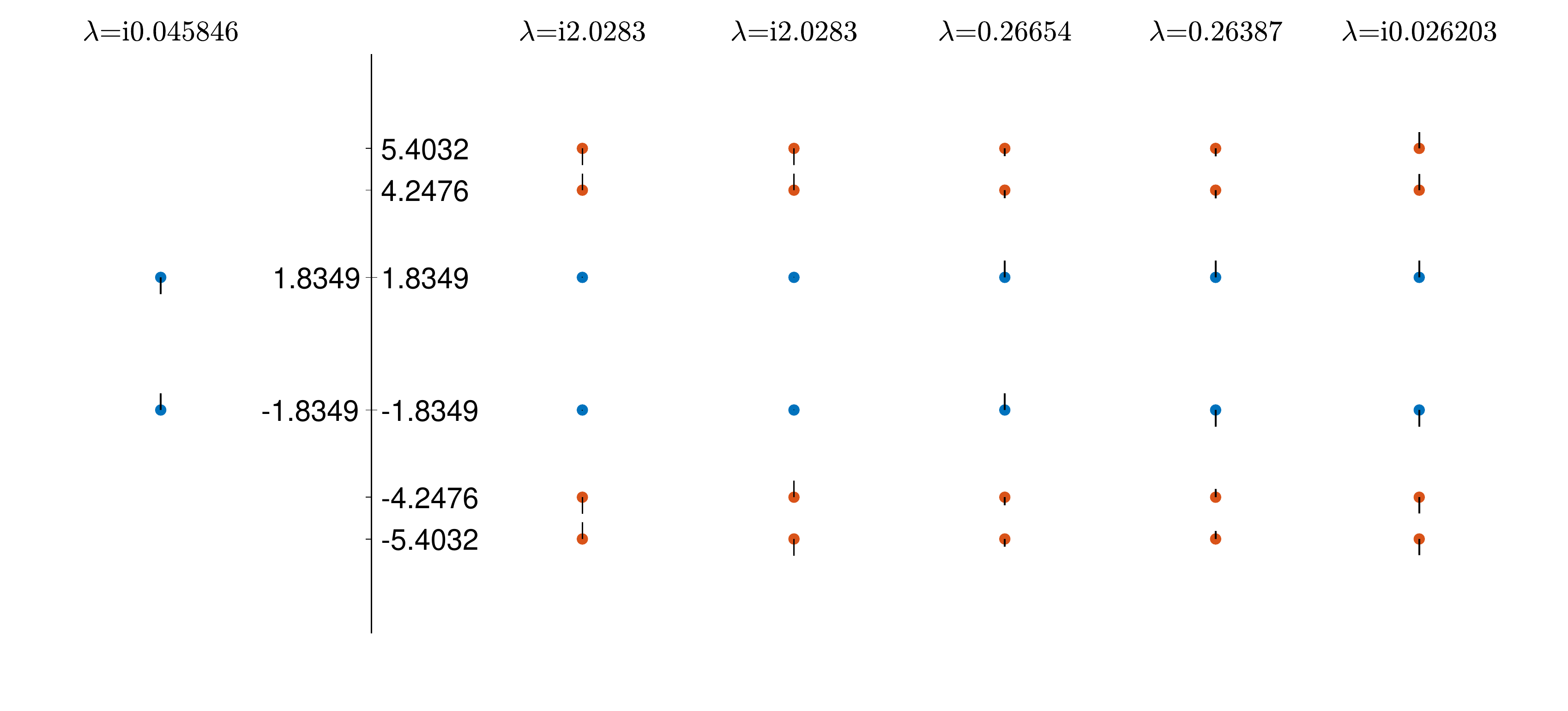}}
\subfigure[  ]{
\includegraphics[scale=0.24]{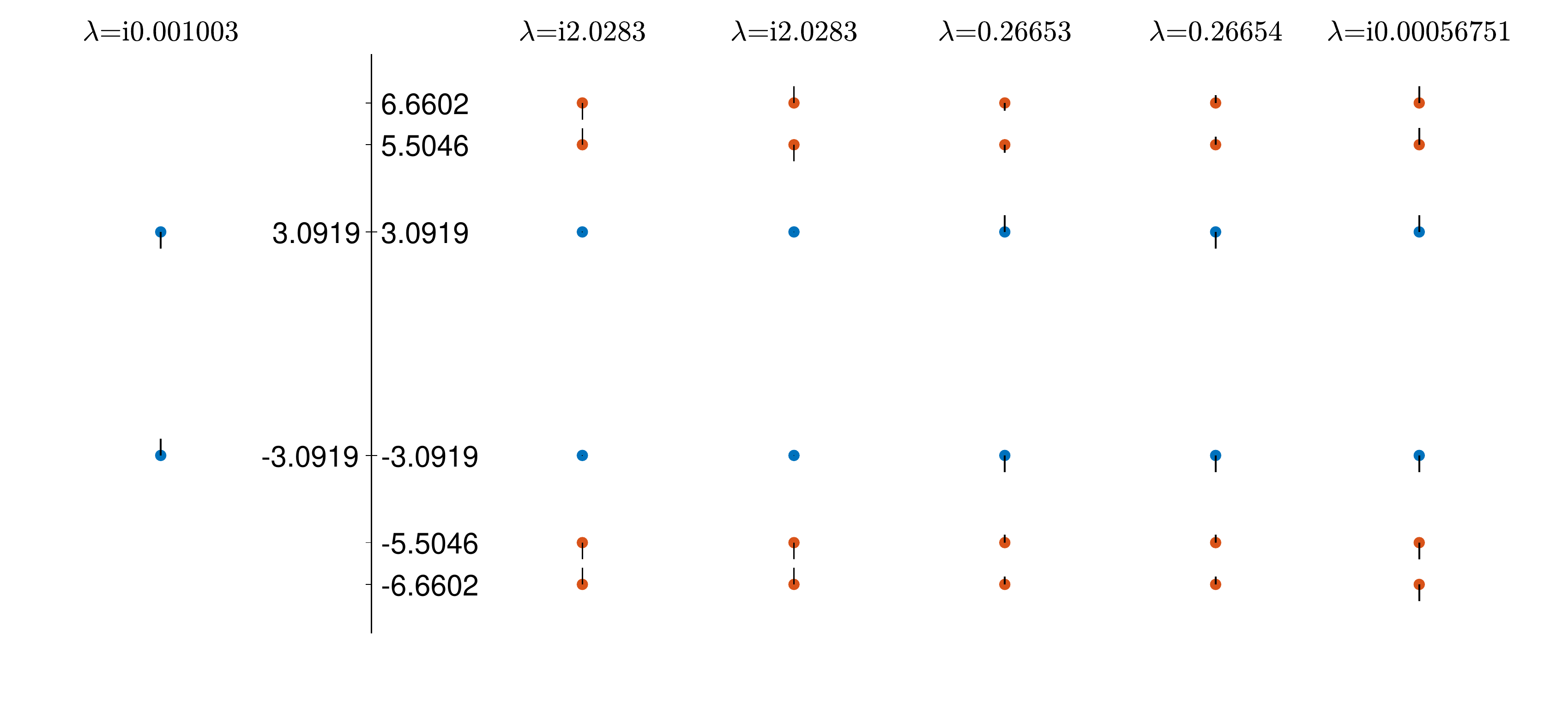}}
\end{center}
\caption{Bifurcation diagrams and the corresponding steady state solutions and spectra for (a) Family 2 (b) Family 4. The former are presented in the format shown previously
of $Q$ vs. $\beta_2$. The latter indicate the eigenvalues and prescribe the motion of
the solitary waves,  in line with the desired eigenmode.}
\label{fig:bif_plots_24}
\end{figure}

Lastly, we briefly refer to families 2 and 4,
showcased in Fig.~\ref{fig:bif_plots_24}. Here, as explained
at the level of the theoretical analysis of the energy
landscape, but also corroborated by related numerics
of the unstable families, the lower branches concern
solutions that are stable. Indeed these are center
configurations (around which the dynamics may orbit,
as a result of the instability of the saddles above).
This is reflected in the stable nature of the two bottom
right subplots in panels (a) and (b) within Fig.~\ref{fig:bif_plots_24}. As before, crossing through
negative $\beta_2$ in both families leads to modulationally
unstable backgrounds with continuous spectrum crossing
through to the real line. Past the turning point, we 
revert to the upper branches for each configuration
which look fairly similar and essentially differ in the
location of the resulting 6 kinks.
Interestingly the inner kinks remain at the same
distance as for the stable lower branch (for each
of the families 2 and 4) and two outer
pairs of kinks are added to the configuration at larger
distances. 

As regards the upper branches of each of these families,
panels (c) and (d) of Fig.~\ref{fig:bif_plots_24} 
reflect the theoretical predictions for their stability. 
In each case there are two unstable modes (rather than
3) which, in fact, have very proximal eigenvalues.
It is for that reason that these two pairs of real
modes cannot be distinguished in the two upper right
insets of panels (a) and (b). Recall that the
oscillatory instabilities are not systematically
considered here given their size dependence. 
The other 3 pairs of nonzero modes of the 6-kink
system are imaginary and are also given in panels
(c) and (d). In either case, the destabilizing
eigenmodes are similar and involve either
an in-phase motion of the inner 2 kinks while
the outer 4 ones are moving in the opposite 
direction or an out-of-phase motion of the inner kinks
which on each ``side'' (i.e., for $x>0$ or 
$x<0$) is opposite to the motion of the outer
kinks. For brevity, we do not present the
dynamical implementation of these cases, 
although a similarly good agreement with the
predictions of the theory has been found in this
case.

Indeed, we elaborate a bit further on the
quantitative aspects of the comparison of
the theory with our numerical computations now.
This comparison can be seen as summarized
in the two extensive tables~\ref{tab:cases2}
and~\ref{tab:cases3}. The former of these
tables offers the comparison of the equilibrium
configurations in the context of the ODE
theoretical approach of section~\ref{eff_part} and the full PDE
results. In the
latter, the zero crossings of configurations
with 2 (all lower branches), 4 (upper branches
of families 0 and 3) and 6 kinks (remaining upper
branches) have been identified and listed.
One can observe a very good agreement between
the two. This only deteriorates a little in the cases
of outermost kinks but is still qualitatively
excellent and even quantitatively satisfactory.

\begin{table}
\begin{tabular}{||c|c|c|c||}
\hline
\hline \multicolumn{1}{||c|}{Family}&\multicolumn{1}{c}{Branch}&\multicolumn{2}{|c||}{Soliton Position} \\ 
& & ODE & PDE   \\ \hline\hline
0&lower & 0.5778  & 0.6084 \\ \hline
&upper & 1.2063 & 1.2229 \\ 
& & 2.3620 & 2.4389 \\ \hline\hline
1&lower & 1.2063 & 1.20534 \\ \hline
&upper & 1.2063 & 1.20493 \\
& & 3.6190 & 3.63302 \\
& & 4.7747 & 4.84924\\ \hline\hline
2&lower & 1.8349 & 1.83539 \\ \hline
&upper & 1.8349 & 1.83495\\
& & 4.2476 & 4.26350\\
& & 5.4032 & 5.47906 \\ \hline\hline
3&lower & 2.4634 & 2.46336\\ \hline
&upper & 0.5778 & 0.60799\\
& & 2.9905 & 3.03715\\ \hline\hline
4&lower & 3.0919 & 3.09138 \\ \hline
&upper & 3.0919 & 3.27728\\
& & 5.5046 & 5.70647\\
& & 6.6602 & 6.92233\\ \hline\hline
5&lower & 3.7204 & 3.72945\\ \hline
&upper & 1.2063 & 1.22286 \\
& & 2.3620 & 2.43858\\
& & 4.7747 & 4.86794\\ \hline\hline
\end{tabular}
\caption{\label{tab:cases2} Soliton positions, ODE versus PDE, for $\beta_2=0.5$}
\end{table}

\begin{table}
\begin{tabular}{||c|c|c|c|c|c||}
\hline \hline 
Family & Branch & \multicolumn {2}{c|}{Eigenvalues} & Soliton Initial & PDE Eigenvector  \\ 
& & ODE & PDE & Directions & Symmetry \\ \hline\hline
0&lower &2.0337 i  &2.0263 i & $ \downarrow \uparrow$ & even  \\
 \hline
&upper & 0.2139 & 0.2231 & $ \downarrow \downarrow \uparrow \uparrow$ & even  \\
& & 2.0337 i & 2.2072 i * & $\uparrow \downarrow \downarrow \uparrow$ & odd \\
& & 2.0227 i & 2.2481 i * & $\downarrow \uparrow \downarrow \uparrow$ & even \\ \hline\hline
1&lower  & 0.30996  &0.30119 &$ \downarrow \uparrow$ & even\\ \hline
 &upper  & 0.39091  &0.38075 &$\downarrow \downarrow \uparrow \downarrow \uparrow \uparrow$ & even\\
&  & 0.26654  &0.26383 &$\uparrow \uparrow \downarrow \downarrow \uparrow \uparrow$ & odd\\
&  & 0.11958  &0.12634 &$\downarrow \downarrow \downarrow \uparrow \uparrow \uparrow$ & even\\ 
&  &2.0284 i  & 2.0334 i  & $\uparrow \downarrow \downarrow \downarrow \downarrow \uparrow$ & odd\\
&  &2.0284 i  & 2.2871 i * &  $\downarrow \uparrow \uparrow \downarrow \downarrow  \uparrow$ & even \\\hline\hline
2&lower&  0.04585 i    &0.04460 i&$ \downarrow \uparrow$ & even \\ \hline
&upper  & 0.26654  & 0.26401& $\uparrow \uparrow \downarrow \downarrow \uparrow \uparrow$& odd\\
& & 0.26387  & 0.26168 & $\downarrow \downarrow \uparrow \downarrow \uparrow \uparrow$ &even\\
& &0.02620 i &0.02721 i &$\downarrow \downarrow \downarrow \uparrow \uparrow \uparrow$ & even\\
& &2.0283 i  &2.0418 i &$\downarrow \uparrow \uparrow \downarrow \downarrow  \uparrow$ & even \\
& &2.0283 i  & 2.2842 i * & $\uparrow \downarrow \downarrow \downarrow \downarrow \uparrow$& odd\\ \hline\hline
3&lower  &   0.006781 &0.00660 &$ \downarrow \uparrow$ & even\\ \hline
&upper  & 0.30542  &0.30825 &$\uparrow \downarrow \downarrow \uparrow$ & odd\\
& & 0.22037  &0.21038 & $ \downarrow \uparrow \downarrow \uparrow$& even\\
& &2.0227 i  & 2.2318 i * & $ \downarrow \downarrow \uparrow \uparrow$ & even \\ \hline\hline
4&lower  &0.001003 i   &0.000976 i &$ \downarrow \uparrow$ & even\\ \hline
&upper  &  0.26653 &0.26407 & $\downarrow \downarrow \uparrow \downarrow \uparrow \uparrow$ & even\\
& & 0.26654  &0.26402 & $\uparrow \uparrow \downarrow \downarrow \uparrow \uparrow$ & odd\\
& &0.00057 i & 0.00021 & $\downarrow \downarrow \downarrow \uparrow \uparrow \uparrow$ & even\\
& &2.0283 i  &2.0367 i & $\downarrow \uparrow \uparrow \downarrow \downarrow  \uparrow$ & odd  \\
& &2.0283 i  & 2.2784 i * & $\uparrow \downarrow \downarrow \downarrow \downarrow \uparrow$ & even \\ \hline\hline
5&lower  & 0.00014  &0.00014 &$ \downarrow \uparrow$ & even\\ \hline
& upper & 0.30393  &0.31054 & $\downarrow \uparrow \uparrow \downarrow \downarrow  \uparrow$ & even\\
& &  0.26654 &0.26397 & $\uparrow \downarrow \downarrow \downarrow \downarrow \uparrow$ & odd\\
& &  0.15465 &0.15087 & $\downarrow \downarrow \downarrow \uparrow \uparrow \uparrow$ & even\\
& &2.0283 i  & 2.2992 i *& $\uparrow \uparrow \downarrow \downarrow \uparrow \uparrow$ & odd\\
& & 2.0171 i & 2.0213 i & $\downarrow \downarrow \uparrow \downarrow \uparrow \uparrow$ & even\\\hline\hline
\end{tabular}
\caption{ODE versus PDE eigenvalues and PDE eigenvector symmetry for $\beta_2=0.5$. The PDE eigenvalues listed are those that were identified using an inverse participation ratio (IPR) plot; * represents eigenvalues that were not apparent from the IPR plot, but were the closest PDE eigenvalues to the corresponding ODE eigenvalues that also had the same initial direction signature. }
\label{tab:cases3}
\end{table}

An even more stringent test of the theory 
(in comparison to equilibrium positions of
ODE vs. PDE) consisted of the examination 
of the relevant internal modes of vibration
presented in Table~\ref{tab:cases3}. Here, we
have included for completeness the motion induced
by the mode (e.g., as we have already discussed,
all lower branch non-vanishing modes
should be out-of-phase), as well as the
spatial parity of the mode. The former
motion is useful towards understanding the unstable
dynamics induced by the mode (this was also
explained in the discussion of the different
families above). The latter is in line with
the expectations of Sturm-Liouville theory, wherever
appropriate (given the 1d nature of our system).
Remarkably, we see that in this case as well,
the effective particle method of Section~\ref{eff_part} is fairly
accurate in its prediction of both the 
oscillatory and the growing modes of the system.
As the table shows, this turns out to be the
case for both lower and upper branches,
and for all the different families
considered from 0 to 5. It is important
to note here that for the real modes,
such a comparison is relatively straightforward
as the modes are separated from the rest of
the spectrum. However, such a comparison
is far more involved when we are, in principle,
seeking localized modes involving 
relative kink motions ``buried'' within the
continuous spectrum. 
Nevertheless, we have developed a technique
based on the inverse participation ratio (IPR)~\cite{IPR}
which enables us to identify modes
with high IPR, even when embedded in the continuous spectrum,
and to compare them favorably in many cases with the theoretical predictions.
We now briefly discuss the associated details.

\begin{figure}[h!]
     \includegraphics[width=0.6\textwidth,height=0.4\textwidth]{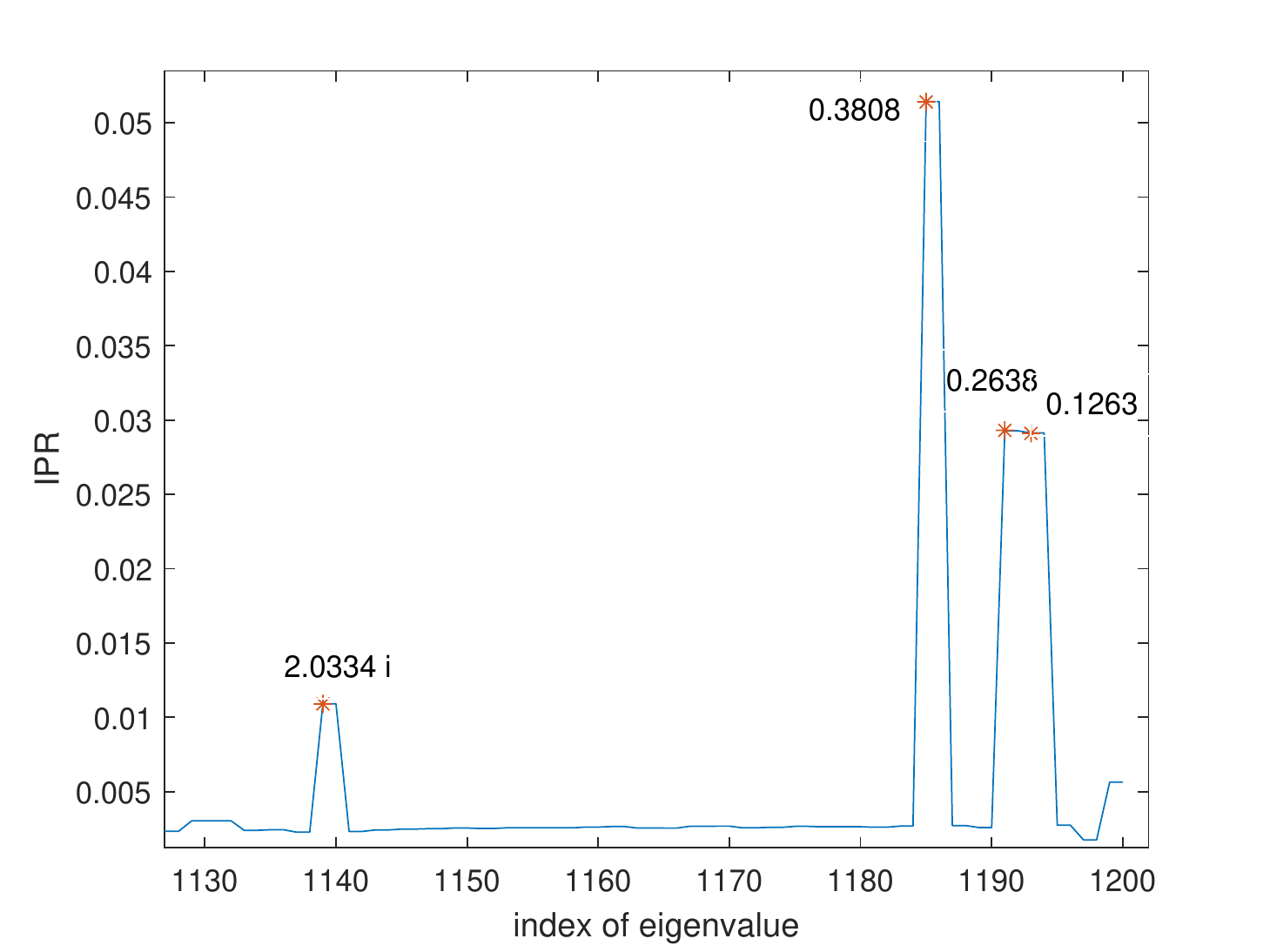}
     \caption{Inverse Participation Ratio plot for Family 1, upper branch. Numerical values shown are eigenvalues corresponding to the eigenvector whose IPR is calculated and plotted. Eigenvectors with index values smaller than shown do not contribute significant IPR values.}
     \label{fig:state1_IPR2}
\end{figure}

The Inverse Participation Ratio can be defined for a function $u(x)$ as
\begin{eqnarray}
\textrm{IPR}= \frac{\int |u|^4 dx}{(\int |u|^2 dx)^2}.
\label{IPR}
\end{eqnarray}
 When $u$ is an eigenvector (eigenfunction), this quantity can be used to find eigenvectors that are the most localized, even when there is a continuous background present. We create an IPR plot, which gives the IPR value for each eigenvector, listed in order of the corresponding eigenvalue (using Matlab's default method of ordering complex eigenvalues). An example, corresponding to Family 1, upper branch, is given in Figure \ref{fig:state1_IPR2}. Since the eigenvalues come in pairs ($\pm$ pairs for the real eigenvalues and complex conjugate pairs ---in fact, quartets $\pm \lambda_r \pm i \lambda_i$--- for the complex valued ones), only the first of each pair that ``stands out'' from the others is marked with an asterisk and labeled with its eigenvalue. From this plot we infer that three real and one purely imaginary eigenvalue correspond to the most localized eigenvectors. Note that the slightly elevated parts of the graph near eigenvalue order number 1130 correspond to eigenvalues that have both non-zero real and imaginary parts (which are not considered) and the elevated part near 1200 corresponds to a zero eigenvalue (representing translational invariance). Thus, the four eigenvalues identified in the figure are the ones listed in Table~\ref{tab:cases3}. Also note that the eigenvalue 2.2871 i listed in Table \ref{tab:cases3} has an asterisk, indicating that it does not correspond to an elevated IPR value.

\begin{figure}[h!]
\subfigure[]{\includegraphics[scale=0.37]{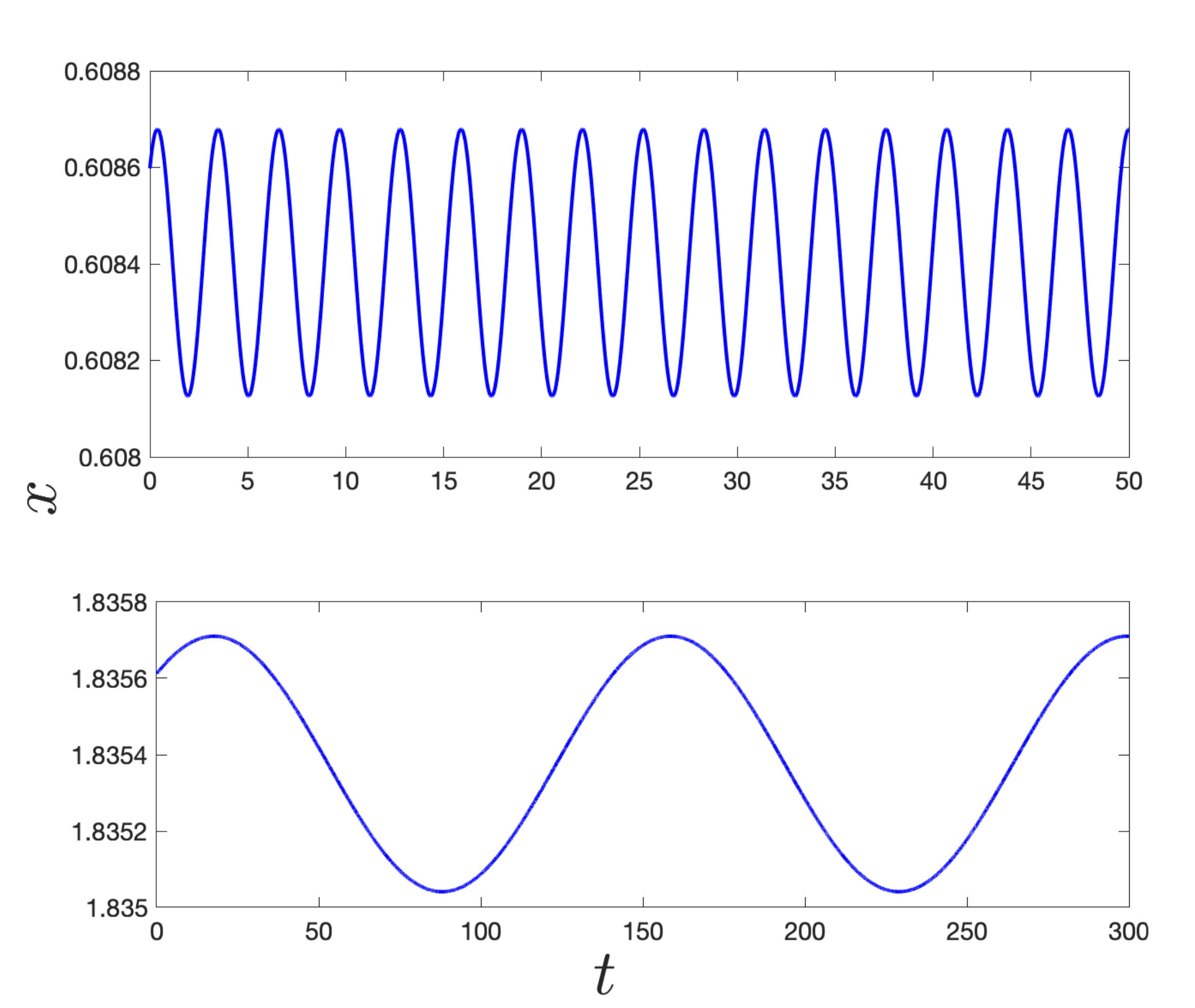}}
 \subfigure[]{    \includegraphics[scale=0.37]{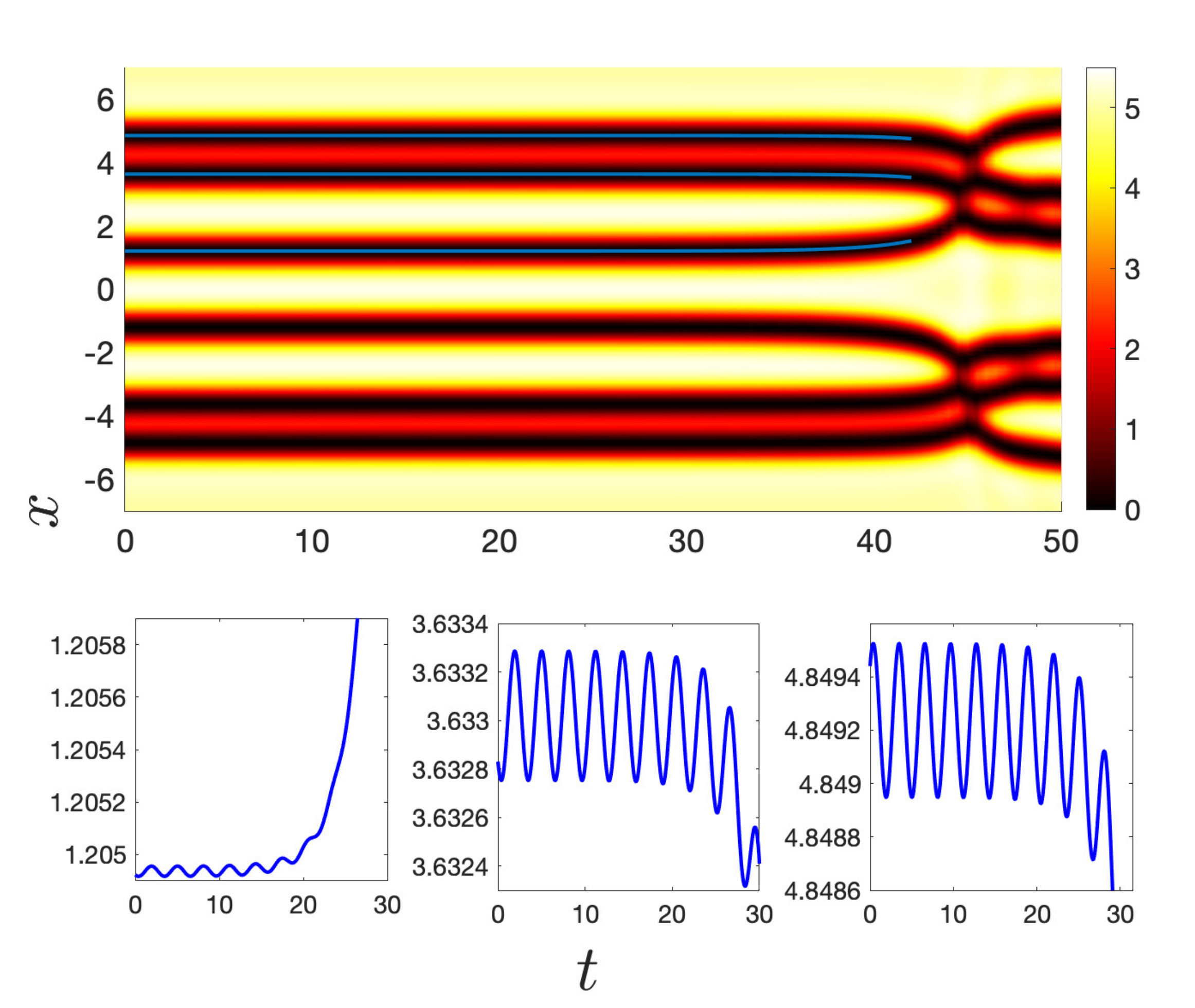}}
     \caption{Dynamics corresponding to imaginary eigenvalues that are embedded in the continuous spectrum. In each case a small amount of an eigenvector with imaginary eigenvalue is added to the steady state, inducing an out-of-phase oscillation for a pair of solitons. The two panels in (a) represent out-of-phase oscillations for Families 0 (top figure, eigenvalue 2.0263) and 2 (bottom figure, eigenvalue 0.0446), both for the bottom branch. We show only the curve that represents the center of the soliton that appears on the positive side of the $x$-axis (and hence on top in the contour plots). All figures in (b) represent Family 1, top branch, with eigenvalue 2.0334. The three blue curves on the bottom again represent the motion of the center of each of the three solitons that appear on the positive side of the $x$-axis (corresponding to the top three solitons in the contour plot shown). These blue curves also appear superimposed on the contour plot, where due to scaling, the oscillations are not apparent.}
     \label{fig:oscillations}
\end{figure}

Figure \ref{fig:oscillations} shows the dynamics for several embedded (purely imaginary) eigenvalues. The plots in the first (left) column verify that for typical lower branch cases, pure oscillations occur for long periods of time, with the frequency given by the corresponding eigenvalue (the top left plot is also, in fact, unchanged  up to $t=300$). The plots in the second column show that for typical upper branch plots, the expected oscillations (with frequencies corresponding to the ---imaginary part of the--- respective eigenvalues) occur for short periods of time, after which nonlinearity takes over as the solitary wave paths start to interact. The blue curves track the centers of the kinks, and are needed as the contour plots do not have fine enough resolution to show the oscillations. The oscillations manifested in these graphs (and their localized
nature around the kink equilibria) confirm that the modes selected by the high
IPR are embedded ones within the continuous spectrum associated with the effective
normal modes of the kink-antikink interacting particle system.

\begin{figure}[h!]
     \includegraphics[width=0.4\textwidth]{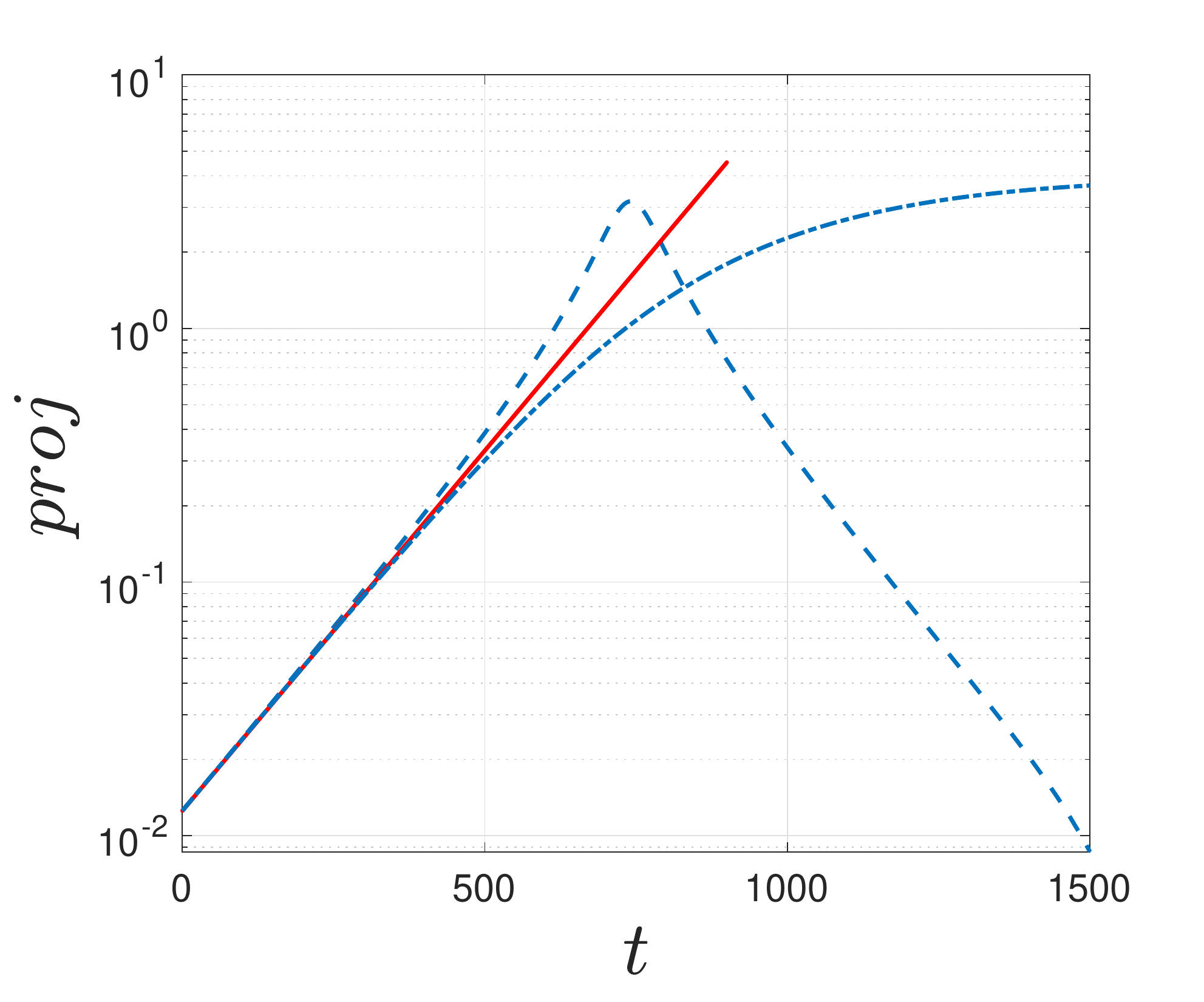}
     \includegraphics[width=0.4\textwidth]{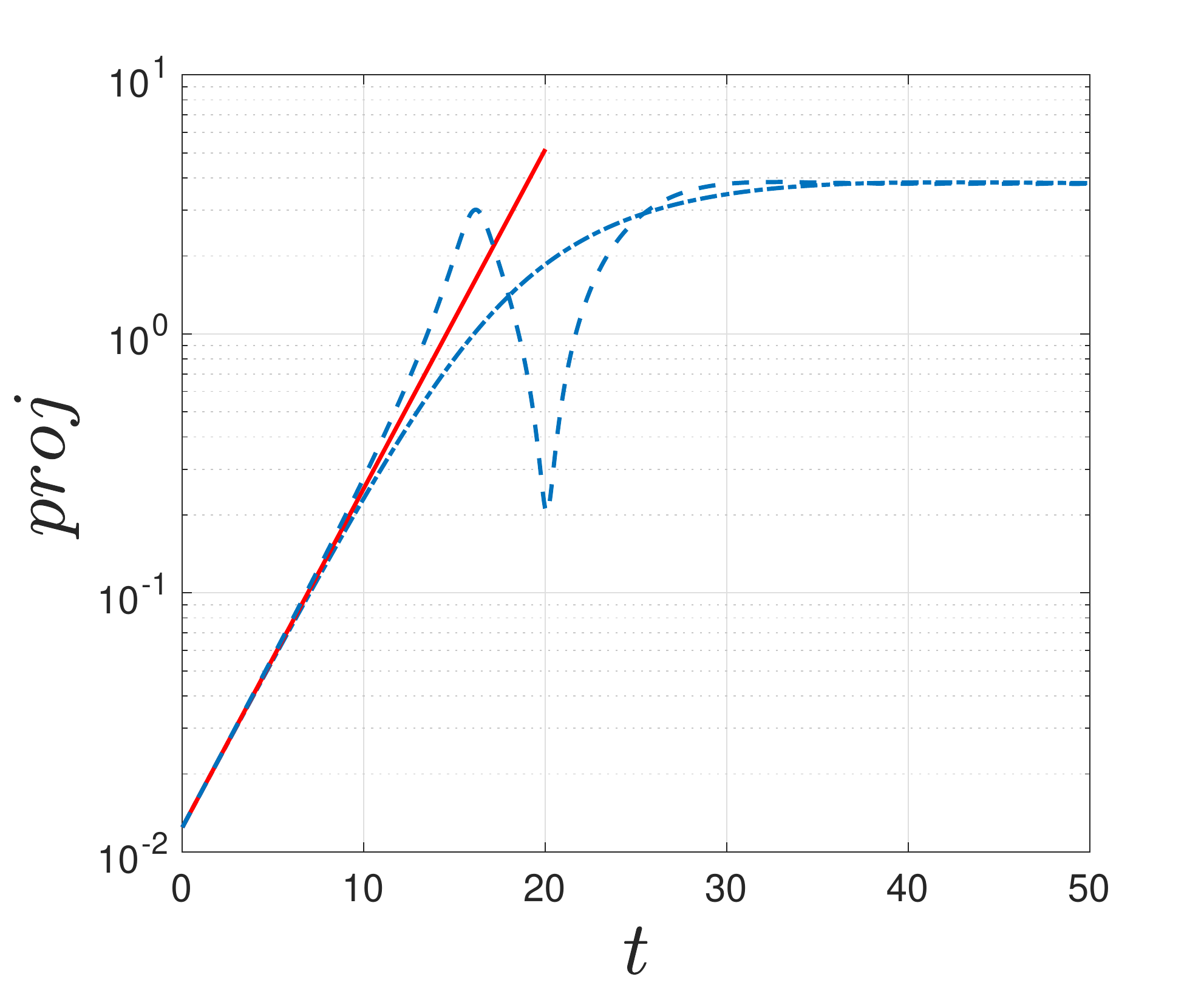}
     \includegraphics[width=0.4\textwidth]{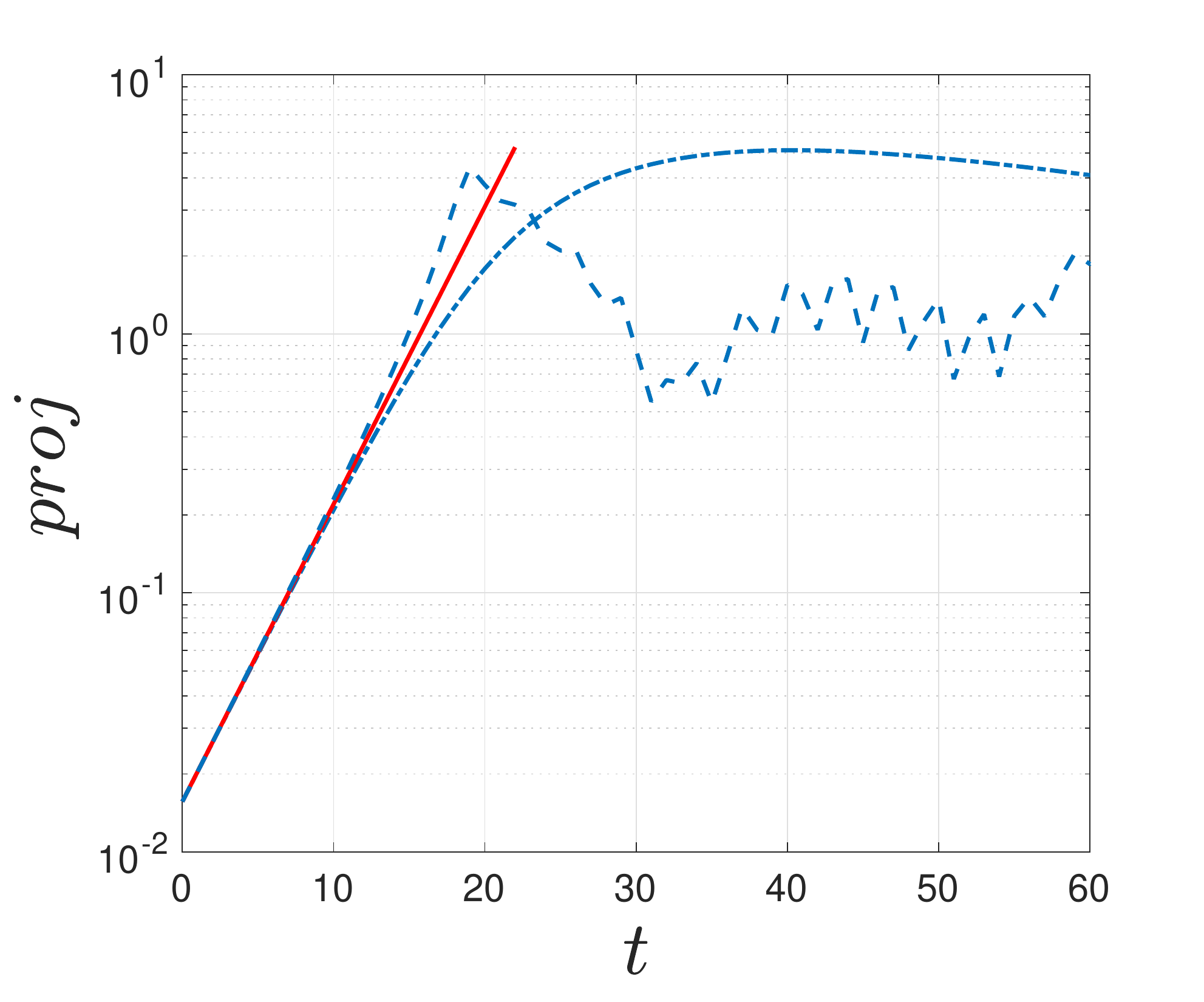}
     \includegraphics[width=0.4\textwidth]{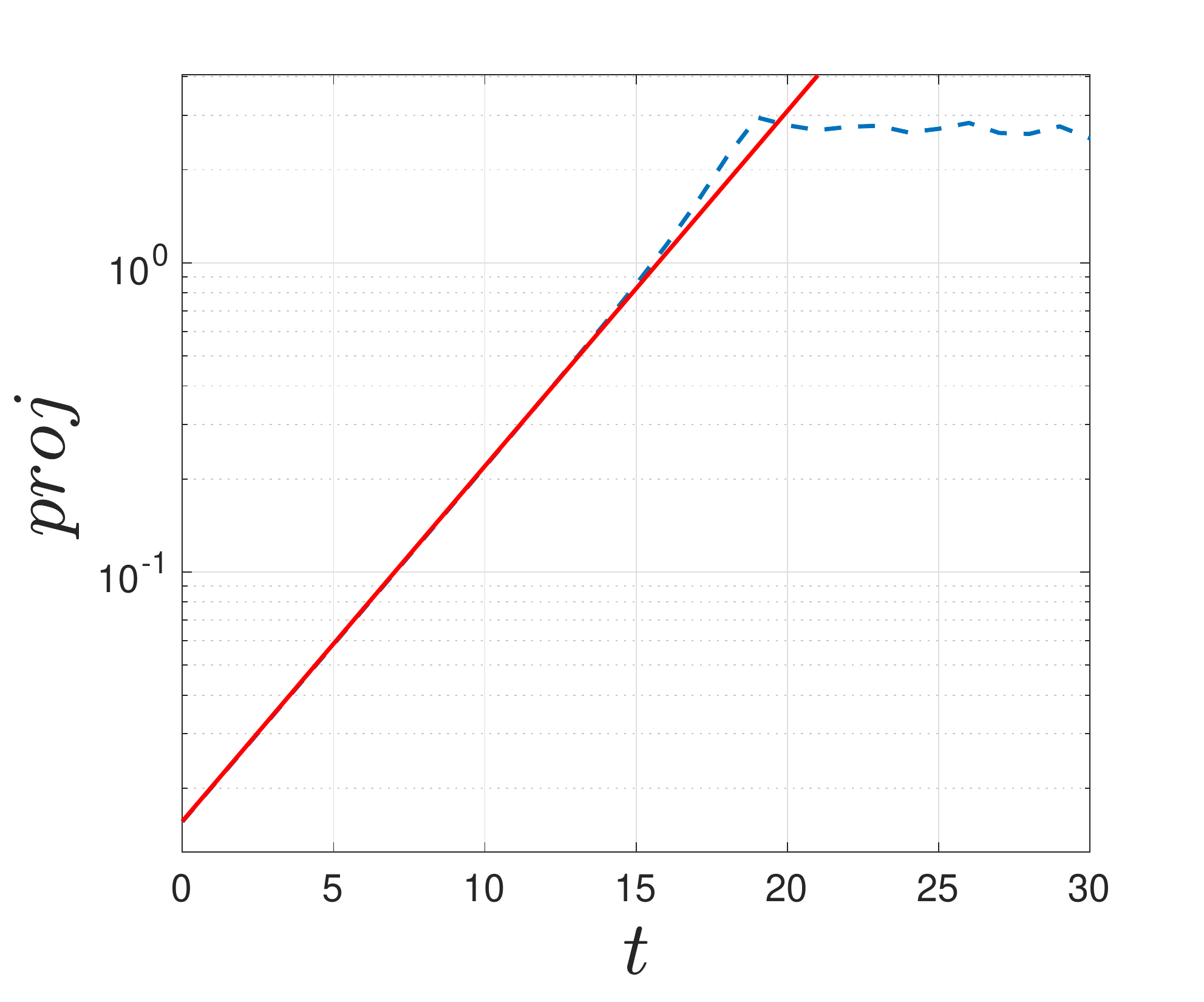}
     \includegraphics[width=0.4\textwidth]{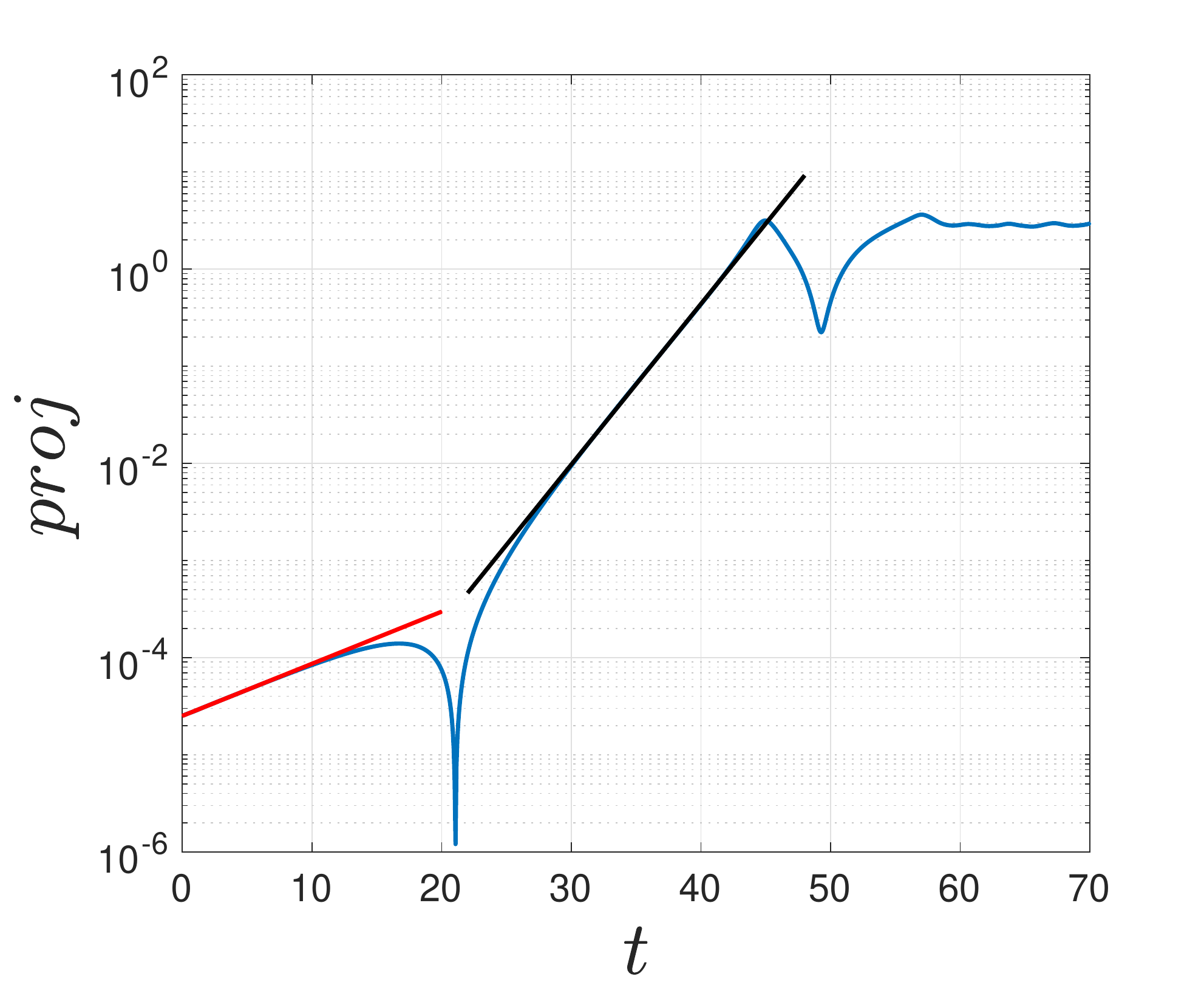}

     \caption{Projection plots for $\beta_2=0.5$. In the top two rows, for a few selected solutions $u(x,t)$,
     we plot the scalar projection of $u(x,t)-u_0(x)$ in the direction of an eigenvector (as a function of time) using a semilog scale. For these solutions, the initial steady state $u_0(x)$ was slightly perturbed in the direction of said eigenvector. In each case the blue lines represent the projections, with the dotted blue lines representing positive perturbations, the dash-dot blue lines representing a negative perturbation;  the red lines represent a least-squares straight line fit to the linear part of the blue curves. We observe a linear portion near the beginning of each plot, whose slope matches very closely with what is predicted by the corresponding eigenvalue. In all cases the slope of the projection curve matches the eigenvalue to two (for the smallest eigenvalues) or three decimal places. The cases are as follows. First row - steady state 3 (left) and steady state 1 (right), both lower branch (note the different time scales). Second row - steady state 4, largest real eigenvalue (left - even eigenvector) and steady state 4, second largest real eigenvalue (right - odd eigenvector - projections coincide). The figure in the third row shows how an initial (small) growth rate can transition to a larger growth rate (projection in blue). This figure corresponds to family 1, upper branch where the initial growth rate of 0.124 (fitted line in red) transitions to a growth rate of 0.381 (fitted line in black). Here $u_0(x)$ was perturbed in the direction of the eigenvector with eigenvalue 0.12634 and then projected onto the eigenvector with eigenvalue 0.38075.}
     \label{fig:projections}
\end{figure}

Lastly, we should mention that in addition
to exploring the growth rate of unstable
configurations via spectral stability analysis,
we have also resorted to an alternative method
to corroborate our numerical stability 
results through
direct numerical simulations. Indeed, we have
considered a method of perturbing the unstable
eigenvectors and subsequently monitoring
the instability growth rates. Typical case examples
of the corresponding
results are shown in Fig.~\ref{fig:projections}.
Here, we compare the findings of the linear stability
computations (via red solid lines) with the
PDE simulations (via blue lines). 
In each case the blue lines represent
the projections arising from subtracting from $u(x,t)$ the equilibrium solution
$u_0$ and then projecting the difference to the instability eigenvector. With the dotted blue lines, we represent the dynamical outcome of positive perturbations, 
while with the dash-dot blue lines the case of a negative
perturbation. The the red lines represent a least-squares straight line fit to the linear part of the blue curves in these semilog plots.
In this way,
we can corroborate the growth rate observed
in the spectral analysis via the instability
dynamics observed in the full PDE model.
In essentially all the cases considered
the agreement is found to be very good with
respect to our theoretical expectations.

\section{Conclusions and Future Challenges}

In the present work we have revisited
the topic of media with competing quadratic
and quartic dispersions in the context
of nonlinear structures commonly considered
in self-defocusing media, namely kink-like
states in the form of dark solitary waves.
We have focused our attention, in particular,
on the setting of multiple such structures
and have proposed a systematic understanding
of pairs of such kinks on the basis of
an energetic landscape emanating from the
kink-antikink interaction. The competition
of the different dispersions, and indeed crucially
the presence of the quartic effects enable the
presence of oscillatory tails and of potential
bound states for multi-kink states.
We have analyzed the first few center-
and saddle-configurations of this
type, indeed 3 center states (families 0, 2, 4)
and 3 saddle ones (families 1, 3, and 5).
In addition to presenting a systematic
continuation of the states in one of the most
natural parametric variations of the system
(the coefficient of the quadratic dispersion), we have
followed the solutions past their (typical,
aside from family 0) turning points, 
identifying their respective upper branches,
unveiling, in turn, solutions associated
with 4, as well as with 6 kinks.

We have provided a systematic particle
picture that offers the possibility of a systematic
classification of the obtained states, irrespective
of the number of kinks based on their interactions,
provided that the kinks are sufficiently well
separated, i.e., for large enough positive 
quadratic dispersion $\beta_2$ in our system.
This analysis was used to accurately capture
the equilibrium distance of the kinks, as well
as their internal excitation modes. In a wide range
of corresponding families and examples, stable
and unstable, lower and upper branch ones,
the method was found to provide systematic
insights regarding the kink dynamics and their stability.

Our considerations offer a systematic
view of the possible stationary multi-soliton
solution families and can be naturally extended
to either higher-order families or heteroclinic
ones involving an odd number of kinks. Both 
directions have been successfully attempted,
although they are not detailed herein.
It should be added here that while in the present
manuscript we have taken an approach that is more mathematically
formal and is driven by our numerical computations, the relevant
considerations are well-positioned to be explored on the basis
of rigorous mathematical theory in the context of the so-called
Lin's method allowing for the consideration of existence of
multi-solitary wave solution~\cite{lin,bjorn}. Such an approach
has recently been extended to kinks in conservative systems
in connection to their existence and stability (in fact, in a discrete
realm) by some of the present authors~\cite{ross}. It would be a very
relevant comparison for the present work if such a method was applied
to the context of the model considered herein.
Yet another  consideration suggested
by our results involves the setting of
traveling excitations. In addition to the loss
of Galilean invariance (in the presence of
quartic dispersion)~\cite{galileo} rendering
interesting the existence and stability analysis
of single traveling kinks, we have found that bound,
breathing states of two kinks are quite common
and would be worth seeking as potentially exact
solutions and to understand their stability.
Such waveforms would be time-periodic in a co-traveling
frame, a feature that would necessitate their consideration
under the prism of Floquet theory.
In an additional dimension of considerations
---pun intended---, the study of coherent
structures in higher-dimensional, such as vortices
in media with competing dispersion operators would
naturally also be a direction of particular interest.
Some of the above studies are currently in progress and
will be presented in future publications.

\vspace{5mm}

{\it Acknowledgements.} This
material is also based upon work supported by the US National
Science Foundation under Grant Nos. DMS-2204702 and PHY-
2110030 (P.G.K.).


\newpage
\begin{thebibliography}{100}

\bibitem{sulem} C. Sulem and P.L. Sulem,  {\it The Nonlinear Schr{\"o}dinger Equation}, Springer-Verlag (New York, 1999).


\bibitem{ablowitz1} M.J. Ablowitz, B. Prinari and A.D. Trubatch, {\it Discrete and Continuous Nonlinear Schr{\"o}dinger Systems}, Cambridge University Press (Cambridge, 2004).


\bibitem{mjarecent} M. J. Ablowitz, {\it Nonlinear Dispersive Waves: Asymptotic Analysis and Solitons}, Cambridge University Press (Cambridge, 2011).


\bibitem{pethick} C.J. Pethick and H. Smith, {\it Bose-Einstein condensation in dilute gases}, Cambridge University Press (Cambridge, 2002).

\bibitem{stringari} L.P.~Pitaevskii and S.~Stringari, {\it Bose-Einstein Condensation}, Oxford University Press (Oxford, 2003).

\bibitem{siambook} P. G. Kevrekidis, D. J. Frantzeskakis, and R. Carretero-Gonz{\'a}lez, {\it The defocusing nonlinear Schr{\"o}dinger equation: from dark solitons and vortices to vortex rings}, SIAM (Philadelphia, 2015).


\bibitem{hasegawa} A. Hasegawa, {\it Solitons in Optical Communications}, Clarendon Press (Oxford, NY 1995).

\bibitem{kivshar} Yu.S. Kivshar and G.P. Agrawal, {\it Optical solitons: from fibers to photonic crystals}, Academic Press (San Diego, 2003).


\bibitem{plasmas} M. Kono and M. M. Skori{\'c}, {\it Nonlinear Physics of Plasmas}, Springer-Verlag (Heidelberg, 2010).

\bibitem{pqs} A. Blanco-Redondo, C. Martijn de Sterke, J.E. Sipe, T.F. Krauss, B.J. Eggleton, and C. Husko, {Nature Communications}, {\bf 7}, 10427 (2016).


\bibitem{pqs3} A.F.J. Runge, D.D. Hudson, K.K.K. Tam, C.M. de Sterke, A. Blanco-Redondo, Nature Photonics {\bf 14}, 492 (2020).


\bibitem{RungePRR2021} A.F.J. Runge, Y.L. Qiang, T.J. Alexander, M.Z. Rafat, D.D. Hudson, A.~Blanco-Redondo, and C.M. de~Sterke, {Phys. Rev. Research} \textbf{3}, 013166 (2021).

\bibitem{pqs2} K.K.K. Tam, T.J. Alexander, A. Blanco-Redondo, and C.M. de Sterke, {Phys. Rev. A} {\bf 101}, 043822 (2020).

\bibitem{bernd} R.I. Bandara, A. Giraldo, N.G.R. Broderick, and B. Krauskopf,
Phys. Rev. A {\bf 103}, 063514 (2021). 

\bibitem{Parker2021} R. Parker and A. Aceves,
Phys. D {\bf 422}, 132890 (2021).

\bibitem{OL22} T.J. Alexander, G. A. Tsolias, A. Demirkaya, Robert J. Decker, C. Martijn de Sterke, and P. G. Kevrekidis, 
Opt. Lett. 47, 1174-1177 (2022).

\bibitem{TsoliasJPA2021} G.A. Tsolias, R.J. Decker, A.~Demirkaya, T.J. Alexander, and P.G. Kevrekidis,  J. Phys. A: Math. Theor. \textbf{54}, 225701 (2021).

\bibitem{OL20} A.F.J. Runge, T.J. Alexander, J. Newton, P.A. Alavandi, D.D. Hudson, A. Blanco-Redondo, and C.M. de Sterke, 
Opt. Lett. {\bf 45}, 3365-3368 (2020). 

\bibitem{galileo} J. Widjaja, E. Kobakhidze, T.R. Cartwright, J.P. Lourdesamy, A.F. J. Runge, 
T.J. Alexander, and C.M. de Sterke
Phys. Rev. A {\bf 104}, 043526 (2021)

\bibitem{KarpmanPLA1994} V.~Karpman, {Phys. Lett. A} \textbf{193}, 355 (1994).

\bibitem{KarpmanPRE96} V.I. Karpman, Phys. Rev. E {\bf 53}, R1336 (1996).

\bibitem{beam_demirkaya} A. Demirkaya and M. Stanislavova, DCDS-B {\bf 24}, 197 (2019).


\bibitem{atanas} I. Posukhovskyi and A. Stefanov, DCDS-A {\bf 40}, 4131 (2020).

\bibitem{atanas2} A. Stefanov, G.A. Tsolias, J. Cuevas-Maraver, P.G. Kevrekidis
J. Phys. A: Math. Theor. {\bf 55} 265701 (2022).


\bibitem{djf}
D.~J. Frantzeskakis, 
{J. Phys. A-Math. Theor.}
  \textbf{43}, 213001 (2010).

\bibitem{manton} N. S. Manton, 
Nuclear Phys. B \textbf{150}, 397 
(1979).

\bibitem{igorb} I.V. Barashenkov,
Phys. Rev. Lett. {\bf 77}, 1193 (1996)


\bibitem{johkiv} M. Johansson, Yu.S. Kivshar, Phys. Rev. Lett. {\bf 82}, 85 (1999).

\bibitem{IPR} B. Kramer and A. MacKinnon,
Rep. Prog. Phys. {\bf 56}, 1469 (1993).

\bibitem{lin} X. B. Lin, 
Proc. Roy. Soc. Edinburgh A {\bf 116}, 295
(1990).

\bibitem{bjorn} B. Sandstede,
Trans. Amer. math. Soc. {\bf 350}, 429
(1998).

\bibitem{ross} R. Parker, P.G. Kevrekidis, A. Aceves,
Nonlinearity {\bf 35}, 1036 (2022).

\end{thebibliography}
\end{document}